\documentstyle[12pt,aaspp4]{article}
\newcommand{\kms}{\mbox{\rm km s$^{-1}$}}          

\begin{document}

\title{DESTRUCTION OF THE GALACTIC \\ GLOBULAR CLUSTER SYSTEM}
\author{Oleg Y. Gnedin \and Jeremiah P. Ostriker}
\affil{Princeton University Observatory, 
       Peyton Hall, Princeton, NJ~08544; \\
       ognedin@astro.princeton.edu, jpo@astro.princeton.edu}

\begin{abstract}
We investigate the dynamical evolution of the Galactic Globular Cluster System
in considerably greater detail than has been done hitherto, finding that
destruction rates are significantly larger than given by previous estimates.
The general scheme (but not the detailed implementation) follows
\cite{AHO:88} (1988; AHO).

For the evolution of individual clusters we use 
a Fokker-Planck code including the most important physical processes
governing the evolution: two-body relaxation,
tidal truncation of clusters, compressive gravitational shocks while clusters
pass through the Galactic disk, and tidal shocks due to passage close to
the bulge. Gravitational shocks are treated comprehensively, using a recent
result by \cite{KO:95} (1995) that the $\langle \Delta E^2 \rangle$ 
shock-induced
relaxation term, driving an additional dispersion of energies, is generally
more important than the usual energy shift term $\langle \Delta E \rangle$.
Various functional forms of the correction factor are adopted to allow
for the adiabatic conservation of stellar actions in a presence of transient
gravitational perturbation.

We use a recent compilation of the globular cluster positional
and structural parameters, and a collection of radial velocity measurements.
Two transverse to the line-of-sight velocity components were assigned
randomly according to the two kinematic models for the cluster system
(following the method of AHO):
one with an isotropic peculiar velocity distribution, corresponding to
the present day cluster population, and the other with the radially-preferred
peculiar velocities, similar to those of the stellar halo.
We use the Ostriker \& Caldwell (1983) and the Bahcall, Schmidt, \& Soneira 
(1983) models for our Galaxy.

For each cluster in our sample we calculated its orbits over a Hubble time,
starting from the {\it present} observed positions and assumed velocities.
Medians of the resulting set of peri- and apogalactic distances and velocities
are used then as an input for the Fokker-Planck code. Evolution of the cluster
is followed up to its total dissolution due to a coherent action of all of
the destruction mechanisms. The rate of destruction is then obtained as 
a median over all the cluster sample, in accord with AHO.

We find that the total destruction rate is much larger than that given
by AHO with more than half of the present clusters ($52\%-58\%$ for the 
OC model, and $75\%-86\%$ for the BSS model) destroyed
in the next Hubble time. 
Alternatively put, the typical time to destruction is comparable to the
typical age, a results that would follow from (but is not required by)
an initially power law distribution of destruction times.
We discuss some implications for a past history
of the Globular Cluster System, and the initial distribution of the destruction
times raising the possibility that the current population is but a very
small fraction of the initial population with the remnants of the destroyed
clusters constituting presently a large fraction of the spheroid (bulge+halo)
stellar population.

\end{abstract}
\keywords{celestial mechanics, stellar dynamics -- globular clusters: general
          -- Galaxy: kinematics and dynamics}

\section{Introduction}
Globular clusters are thought to be the oldest stellar systems in our Galaxy,
and a history of attempts by theoreticians and observers to understand
the keys of their evolution is as old as the discipline of stellar dynamics,
with a comprehensive review provided by \cite{S:87} (1987).
Yet, these relatively simple stellar
systems are not understood well enough to predict their future with desirable
accuracy, in part due to lack of accurate observational values for the current
dynamical state and in part due to a residual uncertainty concerning the
complete catalog of relevant physical processes operating on these systems.
More than that,
we have almost no clues about their past, and in particular, whether what
we see now is representative of the {\it initial} Globular Cluster System, or
just a small leftover after a great destruction battle that occurred 
earlier in the history of the Galaxy.
In this paper we consider the evolution of the globular clusters
and their ultimate disappearance, and propose a simple model
for their initial distribution.

Pre- and post-core collapse evolution of an isolated
cluster is relatively well understood (\cite{S:87} 1987; \cite{G:93} 1993).
Significant progress in understanding of the evolution was
achieved using Monte Carlo and Fokker-Planck calculations.
But the galactic environment makes the clusters subject to external
perturbations -- tidal truncation and the gravitational shocks due to
passages close to the bulge and through the disk.
The shock processes, although known to be important, have never been
carefully included in the evolution of the system.
We investigate the Fokker-Planck models including the shocks
elsewhere (\cite{GLO:96}; hereafter GLO) and show that the dispersion of energy
of the stars, induced by the shocks (\cite{KO:95} 1995; hereafter KO), 
is generally even more 
important for the evolution than the first-order energy shift.
Another very important aspect of modelling globular clusters is the 
initial mass spectrum.
Multi-mass clusters undergo core collapse much faster than in the
single-mass case, and their destruction is much more efficient
(see \cite{LG:95} 1995, and the references in GLO). We restrict ourselves
however to the single mass models in order to maintain clear
physical understanding; but due to omission of this aspect of the problem,
our results provide a lower bound to the rate of destruction of the globular
clusters.

As long as we have (at least, approximate) understanding of the
evolution of a single cluster we can turn to the study of the system
of globular cluster in our Galaxy, and in external galaxies.
\cite{CKS:86} used a semi-analytical Monte-Carlo technique to estimate the
importance of the different mechanisms acting upon the cluster. They 
considered two-body relaxation, tidal stripping of stars,
and the first order tidal shocking effect - due to the crossing disk and
interactions with
giant molecular clouds (GMC). For each of those processes they calculated 
the cluster mass and energy changes associated with them to predict
the evolution.
They followed the cluster evolution only up to core collapse, and assumed
a single-mass King model (\cite{K:66}) for the internal cluster structure.
Tidal heating due to the GMC was found to be negligible compared to the
disk shocks. Note however, that the Galactic model assumed in that work
(\cite{BSS:83}; BSS) is strongly favorable to the disk shock for the clusters
with small orbital radius since the surface density of the disk increases
exponentially as the galactocentric radius
decreases (see Section \ref{sec:GalModel}).
Chernoff et al. (1986) has concluded
that many of the clusters located within inner 3 kpc from the Galactic center
have undergone core collapse, and many of them may already have been
destroyed. A number of authors have pointed out that the bulge and stellar 
spheroid themselves could be composed of remnants of the destroyed globular 
clusters, if prior destruction of globular clusters occurred at high enough
rate. 

Another mechanism for the mass loss is the stellar evolution.
\cite{CS:87} used a similar method to Chernoff et al. (1986)
and included a power-law initial mass
function for the cluster stars. Mass loss due to stellar evolution 
is important for the early evolution of the cluster, but then fades away
because the mass loss is large only for massive
stars whose lifetime is short. 
Comprehensive study including the effects of stellar evolution
has been done by \cite{CW:90}. They used a Fokker-Planck code with an extensive
spectrum of stellar masses (20 species), which includes stellar
evolution and relaxation processes. 
Multi-mass models evolve much faster and the evaporation rate is larger.
Also, mass segregation (\cite{S:87} 1987) speeds up the collapse.
Mass loss during first $5 \times 10^9$ yr is sufficiently strong to
disrupt weakly concentrated clusters ($c<0.6$). Combined with the relaxation,
it destroys many low mass and low concentration clusters within a Hubble
time.

The present characteristics and the evolutionary state of the observed 
Galactic globular clusters were also
investigated by \cite{AHO:88} (1988; hereafter AHO).
We will draw heavily on that paper and compare our results with AHO.
AHO used a sample of the 83 Galactic globular clusters with the known
structural parameters and line-of-sight velocities.
They considered virtually all the important physical mechanisms (except
the mass spectrum and mass loss due to stellar evolution) to calculate 
the present day
destruction rates for the clusters in the sample. The rates were defined
as the inverse time it takes for a given mechanism to dissolve a cluster,
in units of a Hubble time, which is nominally defined as $10^{10}$ yr.
They estimated the rates for
the evaporation through tidal boundary, disk and bulge tidal shocks, 
and dynamical friction.
The last process has not been widely investigated in its application
for the clusters' evolution. Its effect reduces to the gradual spiraling of the
cluster toward the Galactic center, as the cluster loses orbital energy due to
continuous interactions with field stars and dark halo. AHO found that this 
mechanism
is a relatively unimportant one, except for unusually massive clusters.
They calculated a number of orbits
associated with each cluster in the sample and took a median over the whole
resulting distribution for a corresponding destruction rate.
They investigated two Galactic models
(OC and BSS; Section \ref{sec:GalModel}), and two kinematic models for
the globular cluster system: isotropic, which corresponds to the current
distribution, and predominantly radial, which resembles that of the halo stars
and might be closer to an {\it initial} cluster distribution.
The assumption concerning the velocity distribution is necessary as 
only one velocity component is observed.
(Note however that the program to obtain the true space velocities of globular
clusters is underway; see \cite{CH:93} 1993).
We will discuss the kinematic models in Section \ref{sec:kinematics}.

The central bulge was found to be very efficient in destroying clusters
on a highly
elongated orbits. This leads to an {\it isotropization} of the orbits,
and even preferential survival of tangentially biased orbits in the
Hubble time. Thus, an initially radial distribution better fits the present
population, than the initially isotropic one. At the current time, evaporation
is the most important destruction process. The difference between the two
Galactic models was found to be relatively small, except for the disk
 shocking that
obviously depends on a body of the surface mass profile. AHO introduced
a weighting factor that accounts for the fact that some orbits are less
probable because of the various destruction mechanisms acting upon the cluster.
Such a weighting reduces the computed rates, but converges to a similar
value of $0.05$ for all Galactic models and kinematic profiles considered.
This implies that only 4 or 5 clusters are destroyed over a Hubble time
for present conditions. But AHO treated clusters in a simplified fashion
as King models and they (like other investigators) did not allow for the
tidal shock relaxation phenomena. It is these defects that we remedy in the
present paper.

Finally, a recent paper by \cite{HD:92} used a statistical approach and 
approximate analytical
estimates for the relaxation times for a sample of 140 clusters. They concluded
that the current evaporation rate is $5\pm 3~{\rm Gyr}^{-1}$
(about ten times the rate found by AHO), which means that
a significant fraction of the present day clusters would be destroyed
in a next Hubble time.

In this paper we make a significant improvement over AHO by applying
detailed Fokker-Planck calculations
to the real globular cluster sample. We describe the sample in Section
\ref{sec:sample} and the Galactic models in Section 
\ref{sec:GalModel}. Then we discuss the two kinematic models and compare
the resulting 
properties of the orbits for our sample. In section \ref{sec:processes}
we describe the destruction mechanisms involved in the globular cluster
evolution. We conclude section \ref{sec:method} with the history
of our code and the formulation of the numerical strategy.
Section \ref{sec:results} presents our results for all runs.
Finally, we speculate on the possible past history and future fate
of the Galactic globular clusters in Section \ref{sec:discussion}.
Section \ref{sec:conclusions} sums up our conclusions.

\section{Method}
\label{sec:method}
\subsection{Observational Data}
\label{sec:sample}
We have used a recent compilation of globular cluster coordinates
by \cite{DM:93}, and distances and absolute magnitudes from \cite{D:93} (1993).
Cluster concentrations, core radii, and half-mass radii were taken from
\cite{TDK:93}. The core-collapsed clusters were assigned a limiting
concentration of $c=2.5$. We used a constant mass-to-light ratio to
obtain the cluster masses, $(M/L)_V=3$ in solar units
(\cite{D:93} 1993). 

We selected 119 clusters out of 143 with available photometric data.
All clusters in our sample have measured radial velocities which we collected
from several sources. \cite{CH:93} (1993) have derived actual space velocities
for 14 clusters, but we chose not to use those data to maintain homogeneity
of the sample. We assign two unknown velocity components to each cluster
using our statistical method in Section \ref{sec:kinematics}.

The observed parameters of our sample are given in Table 1.
The first two columns are the sequential number and cluster's name. 
The next two
columns are the Galactic coordinates. $D$ is the distance in kiloparsecs 
of the cluster from the Sun, and $R$ is its galactocentric radius.
The seventh column is concentration $c=\log(R_t/R_c)$,
followed by core radius $R_c$ in parsecs.
Tidal radius $R_t$ was calculated from the previous two quantities.
The next column is the cluster mass converted from its luminosity.
The last two columns are the line-of-sight velocities and the reference
numbers to the various sources. The references are listed in the comments
to the Table 1.

\subsection{Galactic Model}
\label{sec:GalModel}
To evaluate the gravitational shocks on the clusters we have used two
models for our Galaxy: the Ostriker-Caldwell (1983) model D-150 (OC), and
the Bahcall, Schmidt, \& Soneira model (1983; BSS). AHO present a detailed
comparison between the two models. In the OC model the disk is represented 
as the difference
of two exponentials which vanishes at the center, whereas BSS's disk density
rises monotonically as $R$ decreases. The BSS model has also a compact nuclear 
component within 1 central kiloparsec. Thus both disk and bulge shocks
(Section \ref{sec:processes}) are expected to be more prominent for BSS
model. As suggested by AHO, we consider the difference of the results
or the two models as a measure of the uncertainty associated with the 
distribution of mass within the Galaxy.

We use the same routine to compute the Galactic model as was done by
AHO (the routine was kindly provided to us by L. Aguilar).

\subsection{Velocity Distribution for the Globular Cluster System}
\label{sec:kinematics}
Kinematic models for the Galactic Globular Cluster System 
(GCS) have been sought widely in the past
(e.g., \cite{FW:80} 1980; \cite{T:89} 1989).
\cite{FW:80} (1980) studied a sample of 66 clusters and found no evidence for 
the radial expansion of the GCS as a whole. Their
best fit to the observed kinematic distribution is that with the 
isotropic velocity dispersion rising with the galactocentric distance 
as $R^{0.2}$. 
An isotropic and isothermal (with constant velocity dispersion) model
is still consistent with their data at the 90\% confidence level.
Later work by \cite{T:89} (1989) included 115 clusters and confirmed the 
absence of the expansion. He found that within the inner 7 kpc to the
Galactic center the velocity distribution is isotropic, but for the outer radii
a velocity ellipsoid with slightly increasing line-of-sight velocity dispersion
may be preferred. Rotation velocity estimates and line-of-sight
velocity dispersions as well as their errors for the both sources are
given in Table \ref{tab:kinematics}.

We repeated the analysis of the velocity distribution using our sample.
We adopted the solar motion relative to the LSR of $(-9,12,7)~\kms$
(\cite{MB:81}), and the circular velocity of the LSR of 220~\kms.
Line of sight projections of the two velocities were subtracted from
the observed radial velocities to obtain ones relative to an inertial frame.
Then the rotation velocity and its uncertainty were estimated using
equations (15) and (16) of \cite{T:89}. After subtraction of the rotation
velocity along the line of sight we end up with the peculiar velocities
with (presumably) zero mean. Both our $v_{\rm rot}$ and $\sigma_{\rm los}$
are in agreement with the above results within errors. Our
adopted values are also summarized in Table \ref{tab:kinematics}.

Following AHO we investigate two {\it initial} kinematic models for
the globular cluster system. The first one has an isotropic peculiar
velocity distribution with constant velocity dispersion. This is the
simplest model still consistent with the observations. We used the
one-dimensional dispersion of $118~\kms$, resulting from our sample.

The second model is anisotropic with velocity ellipsoid axis ratio
at solar position $r_\odot = 8.5$ kpc similar to the
spheroid stellar population II:
\begin{equation}
{\sigma_r^2 \over \sigma_t^2} = 1 + {r^2 \over r_a^2},
\label{eq:sigmas}
\end{equation}
where $\sigma_r$ and $\sigma_t$ are the one-dimensional dispersions
radial and transverse to the Galactic center, respectively.
We have adopted the AHO value of the anisotropy radius
$r_a = 0.8^{1/2} r_\odot \approx 7.6$ kpc. 
Cluster orbits for this distribution are nearly isotropic within $r_a$,
and become more and more radial with increasing distance from the center.

The amplitude of the radial velocity dispersion can be obtained from
Jeans equation for the constant circular velocity potential.
Integrating the Jeans equation in spherical coordinates
gives (e.g., \cite{O:65}):
\begin{equation}
{d \rho \sigma_r^2 \over dr} \, + \, {\rho \over r} \, 
\left(2\sigma_r^2 - 2\sigma_t^2 - v_{\rm rot}^2 \right)
= -\rho \, {v_{\rm circ}^2 \over r}.
\end{equation}
For the velocity ellipsoid given by equation (\ref{eq:sigmas}) and
the power-law density profile $\rho \propto r^{-\alpha}$, we obtain
\begin{equation}
\sigma_r^2 = (v_{\rm circ}^2 - v_{\rm rot}^2) \,
\frac{{r^2 \over \alpha-2} + {r_a^2 \over \alpha}}{r^2+r_a^2}.
\end{equation}
We accept $v_{\rm circ} = 220~\kms$ as the standard value for the circular
velocity.
\cite{T:89} found $\alpha = 3.5$ for his sample of globular clusters.
In this paper we assume $\alpha = 3$, corresponding to the old spheroid
stellar population.

The distribution of peculiar velocities at a given galactocentric 
position $r$ of a cluster is then (AHO eq. [4])
\begin{equation}
f(v_r, v_{t1}, v_{t2}) = A \, \exp \left[-{v_r^2 \over 2\sigma_r^2} -
{v_{t1}^2+v_{t2}^2 \over 2\sigma_t^2} \right],
\end{equation}
where $v_r$ is the radial galactocentric velocity, and $v_{t1}, v_{t2}$
are the two transverse components.

The distribution of orbital eccentricities is shown in Figures \ref{fig:ecc_oc}
and \ref{fig:ecc_bss} for the OC and BSS models, respectively.
Upper panels in both figures show histograms for the anisotropic velocity 
ellipsoids, and the lower ones for the isotropic distribution. Note how the 
number of large eccentricities changes from the former to the latter.

\placefigure{fig:ecc_oc}

\placefigure{fig:ecc_bss}

\subsection{Orbit Integration}
\label{sec:orbits}
In order to model the gravitational shock experienced by a cluster
flying by the galactic center (bulge shock) we need an estimate of a
perigalacticon distance $R_{peri}$ and orbit shape (eccentricity $e$)
for each cluster from our sample. We use the orbit integration routine
described in AHO, which employs a 4th order Runge-Kutta method with
variable time step. The Galactic model extends up to a distance of
250 kpc, and all orbits beyond that point are discarded.

Only 4 out of the needed 6 phase space
parameters required to start the calculation of orbits are 
known from the observations (Table 1): 3 spatial
coordinates and a velocity component along the line of sight. To complement
the two ``missing'' tangential velocities we used a statistical approach
similar to that of AHO. The two velocities were randomly drawn and then
selected using the rejection method (\cite{P:92}) in order for the total
three-dimensional peculiar velocity be consistent with the assumed
kinematic model (section \ref{sec:kinematics}). The systemic rotation velocity
was then added to the peculiar velocity, and the resulting velocity
with respect to the Galactic center was used for the orbit integration.

We have followed trajectories that each cluster makes in 10 billion years.
Perigalactic and apogalactic distances were calculated as medians
from the set of all orbits. Similarly we determine the velocity
of the cluster at the perigalacticon (which we use to calculate the
amplitude of the bulge shock, section \ref{sec:bulge}), and the
vertical velocity component at the point where the cluster crosses the
disk (to estimate the strength of the disk shock, section \ref{sec:disk}).
Thus only one ``mean'' orbit is used to evaluate the gravitational shocks.

In contrast to AHO, we could not follow every orbit of a cluster
because of the nature of the Fokker-Planck calculations.
The code has the time step controlled by the relaxation processes, and
it seems hard to reconcile with the orbit integration procedure.
We plan however to return to this subject in next paper and try to perform
simultaneous Fokker-Planck and orbital calculations which would allow us
to model the evolution of the clusters in a more natural way.

\subsection{Dynamical Processes}
\label{sec:processes}
\subsubsection{Evaporation}
Two-body relaxation leads to the escape of stars approaching the unbound tail
of the cluster velocity distribution (\cite{A:38}; \cite{S:40}). Tidal
truncation due to the Galactic potential accelerates this process.
Much more dramatic effects result from the gravothermal instability, when
the inner part of cluster contracts (core collapse), and the envelope expands.
This, in turn, accelerates the rate of evaporation of stars from the cluster.
A recent review of pre- and post-core-collapse evolution of a tidally truncated
cluster is given by \cite{G:93} (1993).

It is conventional to express the life-time of a cluster in terms of the
half-mass relaxation time (\cite{SH:71})
\begin{equation}
t_{\rm rh} \equiv 0.138 \, {M^{1/2} R_h^{3/2} \over G^{1/2}
             m_* \ln(\Lambda)},
\label{eq:trh}
\end{equation}
where $M$ is the total cluster mass, $R_h$ is the half-mass radius, 
$m_*$ - average stellar mass, and $\ln(\Lambda) = \ln(0.4N)$ - Coulomb
logarithm, $N$ being the number of stars in the cluster.
\cite{H:61} introduced the ``escape probability''
\begin{equation}
\xi_e \equiv - {t_{\rm rh} \over M} \, {dM \over dt}.
\end{equation}
Here $dM/dt$ is the mass-loss due to two-body relaxation. For the self-similar
solutions H\'{e}non found $\xi_e \approx 0.045$. 
See GLO for detailed discussions of the evaporation time scale.

\subsubsection{Disk Shock}
\label{sec:disk}
When a cluster passes through the Galactic disk, it experiences a time-varying
gravitational force pulling the cluster toward the equatorial plane. 
The characteristic
time-scale for this force is the time it takes the cluster to cross two
vertical scale-heights of the disk, $H$:
\begin{equation}
t_{\rm cross} \equiv {2H \over V_z},
\end{equation}
and $V_z$ is the cluster velocity component perpendicular to the Galactic
plane.
Because of the large orbital velocity of the cluster
and the relatively small vertical extent of the disk,
the crossing time is usually
shorter than the orbital period of stars in the outer part of the cluster.
Owing to the short-term nature of the effect, it was called a 
``compressive gravitational shock''
(\cite{OSC:72}). On average, stars gain energy and the cluster binding energy
is reduced. This accelerates the escape of stars from the cluster through
evaporation.

On the other hand, in the central region of the cluster
the effect of the compressive shock is largely damped, since the stars 
move very fast and their orbits become adiabatically invariant.
The impact of the shock is thus a strong function of
the position of a star inside the cluster. Following \cite{S:87}, we define an
{\it adiabatic parameter}
\begin{equation}
x_d \equiv {2 \omega H \over V_z},
\label{eq:adpar}
\end{equation}
where $\omega$ is the angular velocity of stars inside cluster (assuming for
simplicity circular motions), and the subscript $d$ stands for disk;
$x_d$ represents the ratio of the shock duration to the orbital period
of a star, so that for small values of $\omega$ the term ``shock''
is appropriate.

To evaluate quantitatively the effect of the disk shock, several approaches
have been
proposed in the literature. The simplest one, the impulse approximation,
assumes that the shock is so fast that the star does not change its position
in the cluster significantly over the time $t_{cross}$. However, for the 
reason of adiabatic conservation, this approximation highly
overestimates the impact when $x_d \gtrsim  1$. 
A more careful treatment is given by a harmonic
potential approximation (\cite{S:58}), where we assume all stars,
initially at same radial
distance $r$ from the center of the cluster, move around the center with
the same oscillation frequency $\omega$. Then, referring to the equation (25)
of KO, the average energy shift for every star is
\begin{equation}
\langle \Delta E \rangle_{\rm disk} = {2 \, g_m^2 \, r^2 \over 
                                       3 \, V_z^2} \, A_d(x_d),
\label{eq:disk1}
\end{equation}
where $g_m$ is the maximum gravitational acceleration experienced by stars
due to the disk. Here $A_d(x_d)$ is the factor taking into account the 
adiabatic invariants. For the harmonic approximation (e.g. \cite{S:87},
eq. [5-28])
\begin{equation}
A_{d1}(x) = \exp \left( -{x^2 \over 2} \right).
\label{eq:SpCorr}
\end{equation}
We refer to equation (\ref{eq:SpCorr}) as a {\it Spitzer correction}.
However recently \cite{W:94} showed that small perturbations of stellar orbits
can still grow in a system with more than one degree of freedom.
If the system is represented as a combination of multidimensional 
nonlinear oscillators, it is very likely that some of the perturbation
frequencies will be commensurable with the oscillation frequencies of stars.
Then those orbits receive a significant kick from the perturbation, and thus
no longer conserve their actions.
Averaging over whole cluster can give an appreciable change in velocity
and energy.
Such resonances can occur even for an arbitrary small resonant mode.
Consequently, the adiabatic factor $A_d(x)$ is not exponentially small for
large $x$, but rather a power-law. 
The simplest form of the correction can be written as
\begin{equation}
A_d(x) = \left( 1 + {x^2 \over 4} \right)^{-3/2}.
\label{eq:WCorr}
\end{equation}
For large $x \gg 1$, $A_d \propto x^{-3}$.\footnotemark
\footnotetext{This power law index was suggested by S. Tremaine.}
In the following, we call equation (\ref{eq:WCorr}) the {\it Weinberg
correction}. Figure \ref{fig:AdDisk} shows the difference between the
two forms of the adiabatic correction. A much more detailed discussion
of the Weinberg correction can be found in \cite{GHJO:96} (1996).

\placefigure{fig:AdDisk}

Besides the shift in energy of stars $\langle \Delta E \rangle$, a
gravitational shock also induces a quadratic term 
$\langle \Delta E^2 \rangle$, which governs the dispersion over energy
spectrum and thus pushes some loosely bound stars outside the tidal radius,
speeding the disassociation of the cluster. ``Shock-induced relaxation''
was mentioned briefly in \cite{SC:73} (1973; see also Spitzer 1987, 
p. 116\footnotemark).
\footnotetext{We are indebted to L. Spitzer for pointing out this reference
to us.}
Recently KO noted the importance of 
this effect. We refer a reader to that paper, in which it is
shown that this tidal shock relaxation can be in many cases competitive
with two-body relaxation in causing the evolution of a cluster. Like
ordinary relaxation it causes a spread (diffusion) of initially similar orbits
and ultimately will tend to induce core collapse.

According to KO (their eq. [25])
\begin{equation}
\langle \Delta E^2 \rangle_{\rm disk} = {4 \, g_m^2 \, \omega^2 \, r^4 \over
                                         9 \, V_z^2} \, A_{d2}(x),
\label{eq:disk2}
\end{equation}
where in the harmonic approximation
\begin{equation}
A_{d2}(x) = {9 \over 5} \; \exp \left( -{x^2 \over 2} \right).
\label{eq:SpCorr2}
\end{equation}
Note that the correction in this form does not match unity for $x \ll 1$.
Rather it goes a factor 1.8 over the impulse approximation. The harmonic
potential obviously does not apply for the outer parts of the cluster,
so some mismatch is expected. On the other hand, it enhances the shock
in the cluster halo. We have chosen not to modify the Spitzer's formula,
and compare the calculations with those resulting from the Weinberg correction.
The latter has been used in the same form of equation (\ref{eq:WCorr})
for both energy terms.
We have performed calculations with the two forms of the adiabatic correction,
and also a test case in the impulse regime (no correction was applied,
$A_d \equiv 1$).

Disk shocks occur twice during the orbital period of the cluster, so we
define the disk shock time scale at the half-mass radius (neglecting
the adiabatic vcorrections) as
\begin{equation}
t_{\rm disk} \equiv {P_{\rm orb} \over 2} \, \left( {-E \over \Delta E_h}
              \right)
             = {3 \over 8} \, \left( {V_z \over g_m} \right)^2 \, P_{\rm orb}
               \, \omega_h^2, 
\label{eq:tdisk}
\end{equation}
where $P_{\rm orb}$ is the orbital period, $E \approx -0.2GM/R_h$ - 
cluster energy, and the $\Delta E$ and $\omega$ are evaluated at $R_h$.
Analogously, we find that
\begin{equation}
t_{\rm disk,2} \equiv {P_{\rm orb} \over 2} \, \left( {E^2 \over
            \Delta E_h^2}
           \right) = {3 \over 4} \, t_{\rm disk}.
\end{equation}
Both these terms contribute to the destruction of the cluster
(although by quite different processes: $\langle \Delta E\rangle$ enhances
evaporation and $\langle(\Delta E)^2\rangle$ enhances core collapse),
so that the total destruction rate associated with the disk shock may be 
written roughly as
\begin{equation}
\nu_{\rm disk} \equiv {1 \over t_{\rm disk}} \, + \, {1 \over t_{\rm disk,2}}
           = {7 \over 3} \, t_{\rm disk}^{-1}.
\end{equation}

\subsubsection{Bulge Shock}
\label{sec:bulge}
Similar to the compressive disk shock, every globular cluster
experiences a tidal
shock during its passage close to the Galactic center. The massive compact
component at the center of the Galaxy (bulge) induces a strong tidal force on
the cluster near the perigalactic point of the cluster orbit.
The difference between this effect and that from the smooth and steady
tidal field of the Galaxy is 
primarily due to the time dependence of the bulge shock.

The very close effect of the tidal shock induced by giant molecular clouds
was considered by L. Spitzer as early as (1958). The disturbing object was
represented as a point-mass, and cluster orbit near the point of the
closest approach (perigalacticon in our case) is assumed to be a straight path.
Employing the harmonic approximation we find the energy change for each star
due to the bulge shock:
\begin{equation}
\langle \Delta E \rangle_{\rm bulge} = {4 \over 3} \, \left( {G \, M_b 
         \over V_p \, R_p^2} \right)^2 \, r^2 \, 
         A_{b1}(x_b) \, \chi(R_p) \, \lambda(R_p,R_a).
\label{eq:bulge1}
\end{equation}
Here $M_b$ is the bulge mass, $V_p$ is the cluster velocity at the 
perigalacticon $R_p$, and $A_b$ is the corresponding adiabatic correction.
Two new corrections arise as follows.
The distribution of the bulge mass extends up to many kiloparsecs, and for some
clusters it is not a good approximation to consider the bulge as a point-mass.
Rather,
the tidal field exerted by such an extended mass profile will differ
from the point-mass field (given by, e.g., \cite{S:87}). Details of the
calculation of the correction factor $\chi(R_p)$ allowing for that effect
will be given elsewhere (\cite{GH:96}).

We give here only the final expression which we use in our calculations:
\begin{equation}
\chi(R_p) = {1 \over 2} \, \left[ (3J_0-J_1-I_0)^2 +
            (2I_0-I_1-3J_0+J_1)^2 + I_0^2 \right],
\label{eq:masscorr}
\end{equation}
where
\begin{mathletters}
\begin{eqnarray}
I_0(R_p) & \equiv & \int_1^\infty \, m_b(R_p\zeta) \, {d\zeta \over \zeta^2 \,
           (\zeta^2-1)^{1/2}}, \\
I_1(R_p) & \equiv & \int_1^\infty \, \dot{m}_b(R_p\zeta) \, {d\zeta \over 
           \zeta^2 \, (\zeta^2-1)^{1/2}}, \\
J_0(R_p) & \equiv & \int_1^\infty \, m_b(R_p\zeta) \, {d\zeta \over \zeta^4 \,
           (\zeta^2-1)^{1/2}}, \\
J_1(R_p) & \equiv & \int_1^\infty \, \dot{m}_b(R_p\zeta) \, {d\zeta \over 
           \zeta^4 \, (\zeta^2-1)^{1/2}},
\end{eqnarray}
\end{mathletters}
and $m_b(R) \equiv M_b(R)/M_b$ is the normalized bulge mass distribution
at radius $R$, with $\dot{m}_b(R) = dm_b(R)/d\ln{R}$.

\cite{AW:85} (1985; also AHO) proposed a correction of the form
$\chi = {\onehalf} \, I_0^2$ only. Obviously that correction factor 
is everywhere smaller than our's (eq. [\ref{eq:masscorr}]), although 
the difference
is not dramatic. We have plotted both these factors in Figure \ref{fig:MCorr}.
For comparison, we plot also the function $m_b(R)$, which would be the naive
correction which allowed only for the mass interior radius $R$.

\placefigure{fig:MCorr}

The second correction, $\lambda$, is intended to take into account
the time variation of the tidal force along an elliptic orbit of a cluster.
Following AHO, we take the difference of the magnitudes of the tidal force
at perigalacticon and apogalacticon, as the total amplitude of the tidal
effect. Thus the correction factor $\lambda$ can be expressed as
(AHO eq. [11])
\begin{equation}
\lambda(R_p,R_a) = \left[ \, 1 - {M_b(R_a) \over M_b(R_p)} \, 
                   \left( {R_p \over R_a} \right)^3 \, \right]^2,
\end{equation}
where $R_p$ and $R_a$ are the perigalactic and apogalactic distances
of the cluster, respectively.

Finally, $A_{b1}(x)$ is the corresponding adiabatic correction for the bulge
shock. In the harmonic approximation we get the {\it Spitzer correction}:
$A_b(x) = {\onehalf} \, [L_x(x)+L_y(x)+L_z(x)]$ in his notation
(\cite{S:58}, eqs. [36-38]).
For the {\it Weinberg correction} we use again the same function 
(eq. [\ref{eq:WCorr}]) as for the disk shock.
The argument of the $A$ function is now different,
\begin{equation}
x_b = {2 \, \omega \, R_p \over V_p}.
\label{eq:adparb}
\end{equation}
Figure \ref{fig:AdBulge} compares the two corrections.

\placefigure{fig:AdBulge}

As in the case of the disk shock, bulge shock induces the second relaxation 
term
\begin{equation}
\langle \Delta E^2 \rangle_{\rm bulge} = {8 \over 9} \, \left( {G \, M_b 
         \over V_p \, R_p^2} \right)^2 \, \omega^2 \, r^4 \, 
         A_{b2}(x_b) \, \chi(R_p) \, \lambda(R_p,R_a).
\label{eq:bulge2}
\end{equation}
Analogously to the disk case, $A_{b2} = 9/5 A_{b1}$ in the Spitzer regime,
and $A_{b2}=A_{b1}$ for the Weinberg's one.

Similarly we define the bulge shock time scale. In this case however, the
effect occurs only once per cluster orbital period.
\begin{eqnarray}
t_{\rm bulge} & \equiv & P_{\rm orb}\, \left( {-E \over \Delta E_h} \right)
             = {3 \over 8} \, \left( {V_p \, R_p^2 \over G \, M_b} \right)^2 
             \, P_{\rm orb} \, \omega_h^2. \\
t_{\rm bulge,2} & \equiv & P_{\rm orb} \, \left( {E^2 \over \Delta E_h^2}
           \right) = {3 \over 4} \, t_{\rm bulge}.
\end{eqnarray}

Finally, total destruction rate associated with the bulge shock is
\begin{equation}
\nu_{\rm bulge} \equiv {1 \over t_{\rm bulge}} \, + \, {1 \over 
             t_{\rm bulge,2}}
           = {7 \over 3} \, t_{\rm bulge}^{-1}.
\end{equation}

\subsection{Fokker-Planck code}
We calculate the dynamical evolution of the globular clusters from our
sample using an orbit-averaged Fokker-Planck code descended from 
that of Cohn (1979, 1980).
The code has been modified by \cite{LO:87} and 
\cite{LFR:91} to include
a tidal boundary and three-body binary heating. Although the code allows
a multicomponent stellar mass function, we restricted ourselves in this
paper to a single-component case ($m_* = 0.7 \, m_{\sun}$). 
This requires less parametrization of
the numerical models and, we hope, gives clearer understanding of the
physical processes involved. In the future, it certainly will be of
interest to generalize the present calculations, including a realistic
mass function. \cite{LG:95}, for example, considered evaporation of a 
multi-mass cluster in a steady tidal field and found that the mass loss
doubles compare to the single-component case.

Stars beyond the tidal boundary are not lost instantaneously,
but rather follow continuous distribution function $f(E)$, as described
in \cite{LO:87}. This takes into account the fact that the tidal radius
is not a strict ``border'' for the cluster, because the internal force
just balances the Galactic tidal force at that point. Therefore the outer
stars escape only when they go further away from the cluster.
The tidal field of the Galaxy is assumed to be steady and spherically
 symmetric.
The latter is a weakness of a one-dimensional code (for a discussion
see \cite{LG:95} 1995).
The former assumption is valid only for a circular cluster orbit. For the
actual elliptical orbit the maximum tidal stress occurs close to the
perigalactic
point, and probably determines the tidal cutoff radius. A theoretical estimate
for $R_t$ is given by \cite{IHW:83}, who assumed a spherical mass distribution
that increases linearly with galactocentric radius. However, a proper
calculation of the tidal radius is still a challenge for dynamicists,
including its definition itself. For our calculations we used the observed
{\it present day} tidal radii (Table 1), which were obtained by
fitting a single-mass King model to the observed cluster density profiles.

Heating of stars, reversing the core collapse, is due to three-body binaries.
They are included explicitly, without following their actual formation
and evolution, according to the prescription by \cite{C:85}.
Although \cite{O:85} showed that the tidally captured binaries are 
probably more dynamically important for massive clusters than those 
formed through the tree-body interactions, the re-expansion phase following
the core collapse is largely independent of the central heating source
(Henon 1961; Goodman 1993).

\subsection{Philosophy of the Numerical Experiments}
\label{sec:philosophy}
Our goal in this paper is to investigate the importance of the different
destruction processes on the overall evolution of the Galactic Globular
Cluster System. We consider the evaporation process, and the disk and bulge
gravitational shocks. AHO included also the dynamical friction in their
orbit integrations, but we cannot model this effect at the moment. We rely
on the AHO's results that the dynamical friction is not an important
destruction mechanism for most clusters {\it at the present time}. 

Also, AHO evaluated the destruction rates for all of the mechanisms
separately. Using the Fokker-Planck code we could investigate the
effects only together, acting simultaneously. We hope this brings the
numerical simulations closer to the reality, since in the real clusters
all of these processes act coherently, thus ``helping'' each other.
For example, the disk and bulge shock induced relaxation 
(eqs. [\ref{eq:disk2},\ref{eq:bulge2}]) combines with the normal two-body
 relaxation
and enhances the evaporation of stars through the tidal boundary.

Nevertheless, it is of prime interest to rank the processes in their
role for the destruction of the clusters. In particular, AHO
found that the bulge shock is more important mechanism than the disk shock.
Also the newly discovered
tidal shock relaxation (KO) has never been used in the detailed calculations
of globular cluster evolution. How important are these processes?
We try to answer this question using a ``reduction'' approach.
We perform several sets of runs of the Fokker-Planck code,
increasing the number of the processes allowed to act upon the cluster.
The magnitude of each effect can then be estimated by subtracting the
results of the run without that effect from the results of the run
including the effect. Since we do not expect the effects to add in a linear
fashion such an estimate will only approximately determine the strength of the
effect. Thus we organized the following series:
0) evaporation only; evaporation + 1) disk shock, 
$\langle \Delta E \rangle$ term only,
no tidal shock relaxation; 2) disk shock, including the relaxation term
$\langle \Delta E^2 \rangle$;
3) disk shock + bulge shock, without tidal shock relaxation; and finally
4) disk shock + bulge shock, including the relaxation. The last run includes
all the processes we consider in this paper and represents the total
destruction rate computed for the globular clusters.

For each set of runs we repeat the calculations three times, allowing for
different adiabatic conservation factors for the shock processes (sections
\ref{sec:disk} and \ref{sec:bulge}). The first run includes the Spitzer
correction, the second - Weinberg correction, and the last one assumes
the impulse approximation with no adiabatic correction applied.

Each run includes the Fokker-Planck simulations for all of the clusters in our
sample, starting from the present time and ending with their total destruction.
We use the observed concentrations and core radii of the clusters (Table 1)
to model their current structure, which we approximate by a single-mass
King model. The statistically assigned kinematic parameters
($R_p$, $R_a$, $V_z$, $V_p$; section \ref{sec:orbits}) 
were used to estimate the amplitude of the tidal shocks.
We followed evolution of the cluster up to a late stage near total destruction
when the code breaks
down due to numerical difficulties in recomputation the cluster 
potential. The remaining mass at that point is on average 8\% (but no more
than 11\%) of the initial cluster mass. Destruction time $t_d$ was extrapolated
using the least-squares linear fit to the last 10 integration steps. This
gives a more robust estimate of $t_d$ than just linear extrapolation from
the last couple points because many clusters suffer the gravothermal
instability (e.g., \cite{G:93} 1993) at the late stages of their evolution.
We return to this issue in the Results section.

Given the destruction times for all the clusters in the sample, we obtain
the destruction rate in units of a Hubble time
\begin{equation}
\nu \equiv {t_{\rm Hubble} \over t_d} = {10^{10}~{\rm yr} \over t_d}.
\end{equation}
This definition agrees with the destruction rates of AHO, thus allowing
a direct comparison. Since the resulting distribution of the rates is broad
and in general asymmetric around the mean, we choose to take a median 
$\tilde{\nu}$ of the
sample as a characteristic value, in accordance with AHO. Also, a few clusters
from the sample have $\nu \sim 10^3$ and therefore dominate in the mean
and standard deviation. The standard error of the median was estimated as
(\cite{KS:87})
\begin{equation}
\sigma_{\tilde{\nu}} = {1 \over 2 \, N^{1/2} \, f(\tilde{\nu})},
\end{equation}
where $f(\tilde{\nu})$
is the probability density distribution evaluated at
the median point. For our sample we write 
$f(\tilde{\nu}) = \Delta N / N \Delta \nu$, since $f$ should be normalized
to unity. Taking $\Delta N = 0.5N^{1/2}$, we define the two-sided errors
\begin{mathletters}
\begin{eqnarray}
\sigma_{\tilde{\nu}}^+ & = & \nu(N/2 + N^{1/2}/2) - \tilde{\nu}, \\
\sigma_{\tilde{\nu}}^- & = & \nu(N/2 - N^{1/2}/2) - \tilde{\nu},
\end{eqnarray}
\end{mathletters}
for the sorted sample of $\nu$.

\section{Results}
\label{sec:results}
The results of our calculations for the individual clusters are presented
in Table 3. Here we show the destruction rates for the two galactic models
and the isotropic starting kinematics of the clusters (the anisotropic model
is omitted for brevity). The Weinberg adiabatic correction is assumed
throughout the Table. The first two columns identify the clusters and
are the same as in Table 1. The third column gives the half-mass relaxation 
time in years. The remaining columns present the median destruction rates per
Hubble time for each cluster in the sample. The first is the evaporation
rate, followed by the runs including the tidal shocks: first and second order
disk shock, and first order disk+bulge shocks for both the galactic models.
The columns 7 and 10 give the total destruction rates corresponding to our
present time globular clusters. The rest of this section discusses the
statistical results for the sample.

\subsection{Evaporation}
\label{sec:evapres}
We turn first to the evaporation of stars from clusters. 
In other words we perform a set of integrations where we allow only for
ordinary two-body relaxation and losses over the tidal boundary.
Although many studies have been devoted to normal two-body relaxation,
we are unaware of any
complete survey of a distribution of destruction times for different
cluster parameters.
Figure \ref{fig:evap} shows the evaporation time in units of the initial
relaxation time versus cluster structural parameter, concentration
$c = \log(R_t/R_c)$, where $R_t$ and $R_c$ are the tidal and core radii,
respectively. 

\placefigure{fig:evap}

Almost all points on this Figure lie along one curve. 
This makes it very useful 
for quick estimates of the evaporation time given the initial parameters 
of a cluster. The least-squares fit gives
\begin{equation}
{T_{rh} \over T_{ev}} = 1.290\times 10^{-1} - 1.170\times 10^{-1} \, c + 
   3.282\times 10^{-2} \, c^2  + {7.355\times 10^{-4} \over (c-0.55)^2},
\label{eq:fit}
\end{equation}
where $T_{rh}$ is the half-mass relaxation time, and $T_{ev}$ is the
total evaporation time.
Our fit is shown on Figure \ref{fig:evap} as a solid line.

This result suggests that the loosely bound clusters, with $c<0.65$,
are destroyed very fast (in units of the relaxation time). 
The most stable against relaxation are the clusters with $c=1.5-2$, which 
survive for about 40 $t_{rh}$. But this number decreases to 20 $t_{rh}$ 
as the concentration rises beyond 2. Similar behavior has been found by
\cite{J:93} (1993; his Fig. 2).

The correlation between $T_{ev}$ and $T_{rh}$ could be attributed to the
fact that the King models (the initial condition for our calculations) belong
a one-parameter family, and so can be described by the concentration
parameter only. This holds only for the dimensionless quantities, such as
the ratio $T_{ev}/T_{rh}$. The plot of the current relaxation time, expressed
in years (Figure \ref{fig:trh}), does not show any obvious correlation
with the concentration of the clusters.

\subsection{Gravitational Shocks}
Inclusion of the gravitational shocks speeds up the destruction of the clusters
dramatically. Table \ref{tab:results} summarizes the results of our simulations
when all the physical processes are acting on the clusters. The destruction
rates are calculated as medians for the sample 
(see Section \ref{sec:philosophy}). In the OC galactic model the shocks
almost double the destruction rate due to relaxation. Disk shocking 
(``old'' first order) has little effect for both isotropic and anisotropic
kinematic models, but the shock induced relaxation
is more pronounced (0.09 in the median rate gain versus 0.017 in the former
case). The case with the Spitzer correction is enhanced in the shock
relaxation and bulge shocks because of the over-impulse increase of the
adiabatic corrections (Section \ref{sec:disk}).
The bulge shock is much stronger, and in the isotropic model it
completely dominates over the disk shock. For the anisotropic model, the
destruction associated with the bulge shock is reduced. Total destruction rate
for the OC model is in the range $0.45-0.58$.

In the BSS galactic model the disk is stronger, and there is a nuclear
component that dominates the tidal shock over large range of radii
(cf Figure \ref{fig:MCorr}). Therefore the destruction rates are significantly
increased by the shocks: they range from $0.63$ to $0.86$.

Next, we investigate the dependence of the destruction rates on the current
observed cluster position. Figure \ref{fig:ratesr_oc_iso} shows the destruction
rates associated with the several combinations of the physical mechanisms
(Section
\ref{sec:philosophy}) versus galactocentric radius, for the OC galactic model
and isotropic kinematic model for the clusters (Section \ref{sec:kinematics}).
We use here calculations with the Weinberg adiabatic correction.
The left top panel, corresponding to the two-body relaxation, 
is largely determined
by the selection of our sample, since relaxation is an internal process.
However the steady increase in the rate with the decreasing distance to the 
galactic center is most likely explained by the growing tidal field that
imposes tidal cutoff, and therefore removes stars from cluster faster.
Going from the left top panel to the middle bottom we include more and more
tidal shock processes. The disk in the OC model is relatively weak and does not
change the distribution noticeably. Bulge shocks, on the other hand, enhance
the destruction in the center considerably. Relaxation induced by the bulge
shock is comparable to the first order effect. For clusters within 2 kpc
of the center of the Galaxy, tidal shock relaxation dominates ordinary
two-body relaxation in determining the cluster evolution. For such clusters
core collapse occurs much faster and overall evolution proceeds on the
shock time scale, when it is shorter than the relaxation time. During
the cluster contraction, the relative importance of the tidal shock relaxation
to ordinary two-body relaxation decreases and the final stage of core collapse
is described by the self-similar solutions of \cite{H:61}.
After core collapse the
cluster is still so highly concentrated that the density profile is close
to an isothermal sphere. Slow expansion along with the tidal stripping of stars
leads to final dissolution of the cluster.

A similar plot for the anisotropic kinematic model is given in Figure 
\ref{fig:ratesr_oc_ani}.
Figures \ref{fig:ratesr_bss_iso} and \ref{fig:ratesr_bss_ani} show
the distribution for the BSS galactic model and the isotropic and anisotropic
kinematics, respectively.

\placefigure{fig:ratesr_oc_iso}
\placefigure{fig:ratesr_oc_ani}
\placefigure{fig:ratesr_bss_iso}
\placefigure{fig:ratesr_bss_ani}

To emphasize the relative importance of each of the destruction mechanisms,
we define the {\it differential} rates
by subtracting the destruction rates obtained without that process from
the run including the process. Namely, first-order disk shock (``disk1'')
is ``evaporation + disk1'' -- ``evaporation'', second order disk shock
relaxation term (``disk2'') is 
``evaporation + disk'' -- ``evaporation + disk1'',
first order bulge shock (``bulge1'') is 
``evaporation + disk1 + bulge1'' -- ``evaporation + disk1'',
and the bulge relaxation is 
``evaporation + disk + bulge'' -- ``evaporation + disk'' -- ``bulge1''.
Differential rates for the two Galactic and kinematic models
are presented in Figures \ref{fig:ratesd_oc_iso} -- \ref{fig:ratesd_bss_ani}.
For the reasons noted earlier (nonlinearity), these figures must be considered
indicative but not exact.

\placefigure{fig:ratesd_oc_iso}
\placefigure{fig:ratesd_oc_ani}
\placefigure{fig:ratesd_bss_iso}
\placefigure{fig:ratesd_bss_ani}

We present the histograms of the distributions of the destruction rates in
Figures \ref{fig:ratesh_oc_iso} -- \ref{fig:ratesh_bss_ani}.
Arrows at the top show median of the distribution.

\placefigure{fig:ratesh_oc_iso}
\placefigure{fig:ratesh_oc_ani}
\placefigure{fig:ratesh_bss_iso}
\placefigure{fig:ratesh_bss_ani}

We should however be cautious using these differential results, since
all those mechanisms act together and the final result is not a direct
sum of the single processes. For example, relaxation is greatly
enhanced by the tidal shock relaxation and core collapse occurs faster
(in some cases clusters collapsed even when the ordinary relaxation
was not enough to drive the contraction).

\subsection{Vital Diagram for Globular Clusters}
We do not know what was the initial distribution of globular clusters
in our Galaxy. But we can imagine that they occupied some volume in a
given parameter space. All the physical mechanisms considered above
tend to destroy clusters with time, and thus to reduce the allowed volume.
They superimpose particular boundaries which distinguish the present day
clusters from those being already dissolved (or never formed).
\cite{FR:77} considered the cluster mass versus their 
typical size
(half-mass radius $R_h$) diagram. The authors included evaporation, disk shock
and tidal shock heating (for the latter they assume the tidal interaction
with the clusters themselves rather than with the bulge). These processes
cut a triangle on the $R_h - M$ plane, containing the observed
clusters pretty well. 
Ostriker (1975, unpublished) and \cite{CC:84} also included 
dynamical friction, which excludes the very massive clusters.
They noted also that the strength of the destruction mechanisms vary
with galactocentric radius, and therefore the allowed
space depends on the position of a cluster.

We constructed such a vital diagram for the Galactic GCS using our
sample and results of the computations of its evolution. All of the
processes used in the simulations, as well as the dynamical
friction, participate in the diagram. The vital boundary is defined in
such a way, that the sum of all the destruction rates is equal to the
inverse Hubble time:
\begin{equation}
{1 \over t_{\rm Hubble}} = {1 \over t_{ev}} + {1 \over t_{sh}} +
{1 \over t_{df}},
\end{equation}
where $t_{ev}$, $t_{sh}$, and $t_{df}$ are the time-scales over which
a cluster would be destroyed by the given process alone, for the evaporation,
disk and bulge shock combined, and dynamical friction, respectively.
Note that throughout this paper we adopted for simplicity
$t_{\rm Hubble} = 10^{10}$ yr.

According to the calculations reported in Section \ref{sec:evapres}
\begin{equation}
t_{ev} \sim 30 \, t_{\rm rh},
\end{equation}
where $t_{\rm rh}$ is the half-mass relaxation time (eq. [\ref{eq:trh}]).
We plan to perform a similar comprehensive analysis for the gravitational
shocks. At present we use
\begin{equation}
t_{sh} = {1 \over \nu_{\rm disk} + \nu_{\rm bulge}}
\end{equation}
Thus the shock boundary should be considered as a stronger limit.

We have not considered the effects of dynamical friction in details in
this paper. \cite{BT:87} estimated the time for a cluster to lose
its momentum and fall to the Galactic center (their eq. 7-26):
\begin{equation}
t_{df} = {2.64 \times 10^{11} {\rm yr} \over \ln{\Lambda}} \,
         \left( {R \over 2 {\rm kpc}} \right)^2 \,
         \left( {V_c \over 250 {\rm km s}^{-1}} \right) \,
         \left( {10^6 \, M_{\sun} \over M} \right),
\end{equation}
where $\ln{\Lambda}$ is the Coulomb logarithm, $R$ is the {\it initial}
galactocentric distance, $V_c$ is the cluster circular speed, and $M$ is
its mass.

The cluster vital diagram for the OC galactic model
is shown in Figures \ref{fig:diag_oc_iso}, \ref{fig:diag_oc_ani}.

\placefigure{fig:diag_oc_iso}
\placefigure{fig:diag_oc_ani}

The disgram for the BSS model is in Figure \ref{fig:diag_bss_iso},
\ref{fig:diag_bss_ani}.

\placefigure{fig:diag_bss_iso}
\placefigure{fig:diag_bss_ani}

One can see that a significant fraction of the clusters lies outside of the
``surviving'' boundary. This supports our conclusion that many clusters
will be destroyed within the next Hubble time.

\section{Discussion}
\label{sec:discussion}

Using the Fokker-Planck simulations and the sample of the globular clusters
we estimated a current destruction rate per Hubble time. These results are
applicable for the present day clusters evolving forward in time for the
next Hubble time. On the other hand, it is of prime interest to extract 
any possible information regarding the past evolution of the clusters from the
time of their formation up to now. We try to construct a simple model
for the initial distribution of the clusters and to test it against the
current destruction rate. We try then to answer the question raised in the
Introduction: how many of the globular clusters may have been destroyed in 
the history of our Galaxy.

We will not attempt to propose a mechanism for
the formation of globular clusters. An example of a 
possible formation scenario is given by \cite{FR:85}.
Instead, we assume a {\it life-time} function for the globular clusters,
free of any particular assumptions. Let $t_d(t)$ be a time remaining to
the total destruction of a given cluster at epoch $t$. We then define
$f(t_d; t) \, dt_d$ to be the number of clusters with the destruction time 
in the interval $[t_d, t_d+dt_d]$ at time $t$.
Thus, if we treat almost all clusters as made in a short time interval
at the formation of the Galaxy (an obviously gross over-simplification), then
$f(t_d;0) \, dt_d$ gives the initial distribution of cluster lifetimes.
The normalization is
\begin{equation}
N(t) = \int_0^\infty \, f(t_d; t) \, dt_d,
\end{equation}
where $N(t)$ is the number of clusters at epoch $t$.
We would like to advance our function $f$ in time starting from their 
formation, again without a 
detailed prescription for the evolution of the cluster sample.
We simply assume that all the clusters were formed at the same time,
 and that this
time $t_0 \approx 0$ is very small compare to the present Hubble time.
Then number of clusters surviving at the time $t$ is just the number of
initial clusters with $t_d > t$. Thus we have the following relation:
\begin{equation}
f(t_d; t) = f(t_d+t; t_0),
\end{equation}
where we have neglected the contribution of $t_0 \ll t$ in the first argument
of the function $f$.
We call $f(t_d; t_0) \equiv f_i(t_d)$ the initial distribution of the globular
clusters. Integrated over all destruction times $t_d$ it gives the initial
number of globular clusters $N(t=t_0) \equiv N_i$ formed in our Galaxy.

We define also mean $\bar{t}_d$ and median $t_m$ destruction times according to
\begin{equation}
\bar{t}_d(t) = {\int_0^\infty \, t_d \, f(t_d; t) \, dt_d \over
                \int_0^\infty \, f(t_d; t) \, dt_d} =
               {\int_0^\infty \, t_d \, f(t_d; t) \, dt_d \over N(t)},
\end{equation}
\begin{equation}
{1 \over 2} = {\int_0^{t_m} \, f(t_d; t) \, dt_d \over
            \int_0^\infty \, f(t_d; t) \, dt_d} =
           {\int_0^{t_m} \, f(t_d; t) \, dt_d \over N(t)}.
\end{equation}

Now we choose two functional forms for $f_i(t_d)$, which we test
against the distribution of the destruction rates obtained in Section
\ref{sec:results}. The simplest one is that with a constant mean destruction
time for all clusters. It assumes an exponential form
\begin{equation}
f_{i1}(t_d) \, = \, C_1 \, e^{-\alpha t_d}.
\end{equation}
It is easy to show that the mean and median for this distribution are
\begin{eqnarray}
\bar{t}_{d1} & = & {1 \over \alpha}, \\
t_{m1} & = & {\ln{2} \over \alpha}.
\end{eqnarray}
Thus the mean destruction rate does not depend on time, and the evaporation
of the clusters in this case is similar to radioactive decay. Given a current
vital rate we can always calculate how many clusters survived from the
beginning. $N(t)$ goes exponentially to zero
\begin{equation}
N_1(t) = {C_1 \over \alpha} \, e^{-\alpha t}.
\end{equation}

The other function $f$ we consider is a scale free power-law
\begin{equation}
f_{i2}(t_d) \, = \, C_2 \, t_d^{-q}.
\end{equation}
This distribution is not strictly normalizable since the total number of
clusters $N_i$ diverges as $t_0$ approaches zero
\begin{equation}
N_2(t) = {C_2 \over q-1} \, t^{1-q}, \hspace{2cm} q>1.
\end{equation}
However we can successfully apply it to the present time.
The corresponding mean and median are
\begin{eqnarray}
\bar{t}_{d2}(t) & = & {t \over q-2}, \hspace{2cm} q>2, \\
t_{m2}(t) & = & t \, (2^{1 \over q-1}-1).
\end{eqnarray}
Note that both these quantities are proportional to the time of observation.
Thus it seems impossible to determine the initial cluster population for this
distribution given the current rate. Fortunately, we can compare the {\it
shape} of the distribution with the current profile to choose between
the two distributions. In addition we note a very important difference
consequent to the two hypothesized forms for the initial distribution.
In the second (power law) case we should expect that $\bar{t}_d \sim t$;
whenever we look, the time to destruction for the clusters that remain
is of the same order as the age of the existing sample. In the first case
this might occur, but it would require a coincidence between the initial
typical time to destruction and the later point in time when an observer
examines the system. 
{\it The most important point to be made in this paper
is that the numbers in the last line of Table \ref{tab:results} are of
order unity.}
This is consistent with the power law assumption and allows
the possibility of large fractional destruction.

On Figure \ref{fig:vital_oc} we plot a histogram of the destruction rates
for our sample of 119 clusters for the OC galactic model.
The dotted region gives the number of clusters
per logarithmic bin in $t_d$. Two panels correspond to the results
obtained for the isotropic kinematic model with the Spitzer and Weinberg
adiabatic corrections, respectively. We use the medians for the two cases
to determine the shape of the two models for the destruction rates.
The dashed curve corresponds to the exponential model, and the solid line 
is for
the power-law one. Both lines are normalized to the present number of clusters,
$N(t_{\rm Hubble})$.

\placefigure{fig:vital_oc}

Figure \ref{fig:vital_bss} shows the distribution for the BSS galactic model.

\placefigure{fig:vital_bss}

We see from this plot that both models are consistent with the actual
distribution. However, in our view, the power-law case better represents
the overall extended shape of the histogram. The exponential model is more
concentrated around the median value, and falls off very rapidly for
large $t_d$. It predicts also much larger number of clusters in the 
center of the diagram than follows from the Fokker-Planck calculations.
Therefore we favor the second choice, namely, the power-law distribution
of the destruction times. Based on the results obtained for the OC galactic
model and Weinberg adiabatic correction,
power-law index is $q \approx 1.59 - 1.64$ (two limits are for the anisotropic
and isotropic kinematic distributions, respectively).
For the BSS model we find $q \approx 1.73 - 1.82$. 
Thus we conclude that the power-law model with
$q=1.6 - 1.8$ gives naturally the present distribution of the globular cluster
destruction times.

From the last sentence immediately follows that the initial cluster population
could have been much larger than the present one. The present day
characteristics
do not provide a definite number, though they do prefer this scenario.
Thus it is quite possible that a very significant fraction of the original
globular clusters have been already destroyed. Their remnants might constitute
now the inner spherical stellar component of our Galaxy, i.e. the bulge. 
We note also from
Figures \ref{fig:ratesr_oc_iso} and \ref{fig:ratesr_bss_iso} that clusters 
close to the Galactic centers are most
heavily attacked by the bulge shock, and therefore most of the dissolved
clusters could be in the very central region of the Galaxy. A strong
bulge component present in the Ostriker \& Caldwell galactic model
could be enhanced by the input from the destroyed clusters.
Note, that we do {\it not} require the existence of the bulge from the first
days of the Galactic history in order to destroy the clusters. As follows
from the power-law distribution, most of the initially formed clusters
had relatively short life-times, and presumably were not very massive. Thus
the main mechanism driving their dissolution could have been 
internal relaxation.
Though, if present in some form, the central Galactic component would
be very efficient in the cluster destruction.
In any case the overall potential of the Galaxy is changed only very slightly
if we were to imagine that all bulge and spheroid stars were put back into
globular clusters with appropriate orbits. Thus even the bulge shock
tidal evaporation processes would be essentially unchanged, if we were
to put all stars in the quasispherical distribution back into clusters.

Observational data on stellar populations indicate that the spheroid of the
Galaxy is kinematically and chemically distinct from the disk
(\cite{NR:91}). Halo stars are old  (with an age slightly exceeding our adopted
Hubble time of $10^{10}$ yr) and metal poor. Because the density of the
protogalactic gas cloud is not high enough to form stars efficiently,
the oldest stars in the Galaxy are likely to have been formed in
 clusters perhaps not too different from the present population
of globular clusters.
Now we see most of the metal poor stars in the halo, which suggest that
most of them could be left over from the disrupted clusters.
\cite{H:91} proposed two arguments against this idea: 1) orbits of the present
globular clusters are more isotropically distributed than those of spheroid
stars; 2) the clusters have systematically lower metallicity than the field
stars, and become more metal poor with increasing galactocentric distance $R$.
But AHO and \cite{LG:95} (1995) pointed out a possible solution for the 
former problem. The first of these arguments can be
moderated if we consider the initial cluster distribution. AHO argued
that the kinematic model better describing the current population is the
isotropic one (see Section \ref{sec:kinematics}) with the orbits getting
more and more radial as $R$ increases. The destruction mechanisms are the
strongest in the central few kiloparsecs, especially the bulge shock.
Thus the clusters in the most elongated orbits with
small perigalactic distance $R_p$ would be destroyed first. Their remnants
would populate the spheroid as wee see it now, with the preferentially
radial orbits of stars. Surviving clusters, on the other hand, have more
isotropic orbits consistent with the observations. 

To resolve the problem of metallicities, we refer to \cite{vdB:95},
who found that [Fe/H] correlates somewhat more strongly with perigalacticon
$R_p$ than with their current position $R$. Thus the metal rich clusters 
with smaller $R_p$, which would be destroyed first, would enrich the halo
stellar population. Another solution comes from the observed age difference
in the cluster population. The formation period could have been extended up to
$2-5$ Gyr (see references in \cite{NR:91}). Thus the very first low mass
clusters could dissolve due to the internal relaxation, but their stars
would enrich the primordial gas for
the next generation of clusters. Our power-law hypothesis for the distribution
of the clusters predicts that the mean (and median) destruction time is
proportional to the time elapsed since the formation, so that newly formed
clusters would be again most susceptible to disruption. If they were destroyed
within the remaining $7-10$ Gyr, their stars are the population II halo stars
that we see in our Galaxy.

\section{Conclusions}
\label{sec:conclusions}
We have used the Fokker-Planck code to investigate the destruction rate of
globular clusters in our Galaxy. We applied two forms of the adiabatic
correction for gravitational shocks and found that the median results
do not depend much of the particular form of the correction. The current
destruction rate for the sample is about $0.5-0.9$ per $10^{10}$ yr 
(depending on the Galactic and kinematic models), which
implies that more than half of the {\it present} clusters is to be destroyed 
within
the next Hubble time. This estimate is approximately a factor of ten
higher than that obtained by AHO. There are two principal reasons
for the change. First, our Fokker-Planck detailed calculations for each cluster
give systematically larger rates of two body relaxation, and core collapse
than did the essentially time scale arguments of AHO. Secondly, the new tidal
shock relaxation process described by KO further reduces lifetimes by
a significant amount. 

Trying to understand the {\it original} population of Galactic globular
clusters, we considered two possible models for the distribution
of the cluster lifetimes.
Both of them can be normalized to the median present destruction rate.
We favor the power-law model on basis of a shape of the rate distribution
as it naturally explains the fact that the current median time to
destruction is comparable to the present mean cluster age and also because
the predicted distribution of cluster destruction times provides a reasonable
match to observations.
The power-law distribution allows a much larger number of the clusters
to have been formed initially than is currently observed,
and allows the possibility that the debris of the 
early disrupted clusters might have formed the much of spheroid of our Galaxy.

\cite{Su:95} has investigated the possibility of populating the stellar halo
by remnants of the destroyed star cluster (open and globular). He comes to the
conclusion that much larger number of low-mass ($10^3 - 10^4 M_{\sun}$)
clusters than observed now
is required to match the mass of the spheroid. This conforms to our
result, since low mass (and hence, weakly concentrated) clusters have short
lifetime. All those non-observed clusters have to be destroyed before 
the present time.

Finally we note that the inclusion of the mass spectrum in the Fokker-Planck
models would strongly enhance the relaxation and core collapse, and 
ultimately speed up the dissolution of the clusters. This could also
increase the destruction rate by a significant amount.

The destruction rates for the individual clusters are available electronically
upon request.

\acknowledgements
We are greatly indebted to Hyung Mok Lee, who provided us with his
Fokker-Planck code. We thank Luis Aguilar for sharing the orbit integration
routine used by AHO. Many people contributed to the compilation of radial
velocities for our sample. George Djorgovski has been an invaluable source
of data tables and references. We are grateful to Taft Armandroff,
Kyle Cudworth, Chris Kochanek, and Ruth Peterson for their data, and to
Bruce Carney, Chigurapati Murali, Dave Lathau, and Charles Peterson for useful
references. Konrad Kuijken has provided us with the vertical gravitational
force due to the Galactic disk. Finally, we have benefited from the discussions
with Jeremy Goodman, Lars Hernquist, Tomislav Kundi{\'c}, Hyung Mok Lee, 
David Spergel, Lyman Spitzer,
Scott Tremaine, and Martin Weinberg. Most of the calculations have been done
on the SP2 supercomputer at the Maui High Performance Computing Center,
which we greatfully acknowledge. This work was supported in part by
NSF grant AST-9424416.

\setcounter{table}{1}
\begin{deluxetable}{lcc}
\tablecaption{Kinematic parameters of Galactic Globular Cluster System
              \label{tab:kinematics}}
\tablecolumns{3}
\tablewidth{360pt}
\tablehead{\colhead{Reference} & \colhead{${\rm v}_{\rm rot}$} & 
            \colhead{$\sigma_{\rm los}$} \\
            \colhead{} & \colhead{(\kms)} & \colhead{(\kms)}}
\startdata
Frenk \& White (1980) & $60 \pm 26$ & $118 \pm 20$\tablenotemark{a} \nl
Thomas (1989)         & $65 \pm 18$ & $110 \pm 7$ \nl
This paper            & $60 \pm 21$ & $119 \pm 39$ \nl
\enddata
\tablenotetext{a}{The uncertainty of the velocity dispersion is taken
as an average over radial bins of Table 1 from \cite{WF:83}.}
\end{deluxetable}

\setcounter{table}{3}
\begin{deluxetable}{lccccc}
\tablewidth{500pt}
\tablecaption{Median Destruction Rates\label{tab:results}}
\tablecolumns{6}
\tablehead{\colhead{} & \multicolumn{3}{c}{Isotropic} & 
           \multicolumn{2}{c}{Anisotropic} \\
           \cline{2-4} \cline{5-6}
           \colhead{Destruction Process} & \colhead{Spitzer} & 
           \colhead{Weinberg} & \colhead{No correction} & \colhead{Spitzer} &
           \colhead{Weinberg} \\ \cline{1-6}
           \multicolumn{6}{c}{Ostriker \& Caldwell model}}
\startdata
Evaporation & $0.290^{~+0.020}_{~-0.013}$
            & $0.290^{~+0.020}_{~-0.013}$
            & $0.290^{~+0.020}_{~-0.013}$
            & $0.290^{~+0.020}_{~-0.013}$
            & $0.290^{~+0.020}_{~-0.013}$ \nl

+ Disk 1st  & $0.307^{~+0.073}_{~-0.017}$     
            & $0.311^{~+0.069}_{~-0.026}$
            & $0.312^{~+0.069}_{~-0.021}$ 
            & $0.307^{~+0.073}_{~-0.018}$ 
            & $0.309^{~+0.065}_{~-0.017}$ \nl

+ Disk      & $0.397^{~+0.021}_{~-0.086}$
            & $0.391^{~+0.012}_{~-0.080}$
            & $0.394^{~+0.011}_{~-0.049}$
            & $0.393^{~+0.021}_{~-0.074}$ 
            & $0.365^{~+0.033}_{~-0.049}$ \nl

+ Disk 1st + Bulge 1st 
            & $0.467^{~+0.073}_{~-0.058}$ 
            & $0.455^{~+0.074}_{~-0.059}$
            & $0.601^{~+0.096}_{~-0.091}$
            & $0.398^{~+0.088}_{~-0.085}$ 
            & $0.397^{~+0.075}_{~-0.077}$ \nl

+ Disk + Bulge 
            & $0.583^{~+0.133}_{~-0.053}$
            & $0.518^{~+0.108}_{~-0.039}$
            & $0.905^{~+0.048}_{~-0.183}$
            & $0.490^{~+0.138}_{~-0.091}$ 
            & $0.453^{~+0.116}_{~-0.056}$ \nl
\cutinhead{Bahcall, Schmidt \& Soneira model}

Evaporation & $0.290^{~+0.020}_{~-0.013}$
            & $0.290^{~+0.020}_{~-0.013}$
            & $0.290^{~+0.020}_{~-0.013}$
            & $0.290^{~+0.020}_{~-0.013}$
            & $0.290^{~+0.020}_{~-0.013}$ \nl

+ Disk 1st  & $0.495^{~+0.053}_{~-0.088}$
            & $0.498^{~+0.081}_{~-0.075}$
            & $0.502^{~+0.118}_{~-0.027}$
            & $0.395^{~+0.059}_{~-0.057}$
            & $0.395^{~+0.066}_{~-0.049}$ \nl

+ Disk      & $0.654^{~+0.077}_{~-0.101}$
            & $0.660^{~+0.053}_{~-0.124}$
            & $0.818^{~+0.072}_{~-0.159}$
            & $0.511^{~+0.117}_{~-0.045}$
            & $0.493^{~+0.054}_{~-0.058}$ \nl

+ Disk 1st + Bulge 1st 
            & $0.627^{~+0.164}_{~-0.130}$
            & $0.628^{~+0.171}_{~-0.145}$
            & $1.026^{~+0.383}_{~-0.233}$
            & $0.520^{~+0.061}_{~-0.054}$
            & $0.480^{~+0.057}_{~-0.069}$ \nl

+ Disk + Bulge 
            & $0.863^{~+0.163}_{~-0.181}$
            & $0.752^{~+0.276}_{~-0.110}$
            & $1.786^{~+0.747}_{~-0.699}$
            & $0.712^{~+0.103}_{~-0.061}$
            & $0.633^{~+0.072}_{~-0.053}$ \nl
\enddata
\tablecomments{Rates are in fraction of sample per Hubble time
               ($10^{10}$ yr). Total number of clusters in sample: 119.}
\end{deluxetable}


\newpage
\centerline{\bf FIGURE CAPTIONS}

\figcaption[ecc_oc.ps]{Histogram of the orbit eccentricity distribution
  in the OC galactic model,
  for the isotropic and anisotropic kinematic models.\label{fig:ecc_oc}}

\figcaption[ecc_bss.ps]{Histogram of the orbit eccentricity distribution
  in the BSS galactic model,
  for the isotropic and anisotropic kinematic models.\label{fig:ecc_bss}}

\figcaption[disk_ad.ps]{Comparison of the Spitzer 
  [eq. (\protect\ref{eq:SpCorr})] and 
  Weinberg [eq. (\protect\ref{eq:WCorr})] adiabatic corrections
  for the disk shock, i. e., reductions in the energy change due to the shock
  because of the conservation of adiabatic invariants of stellar orbits
  inside the clusters, versus adiabatic parameter 
  [eq. (\protect\ref{eq:adpar})].\label{fig:AdDisk}}

\figcaption[mbulge.ps]{Correction factors versus perigalactic distance
  for the bulge shock energy change due to the extended mass distribution 
  for the bulge. Our correction [eq. (\protect\ref{eq:masscorr})]
  is shown by solid curve, and that from Aguilar \& White (1985), and 
  subsequently AHO, is given by dots. Triangles show the normalized bulge
  mass profile for comparison. If the bulge were treated as a point mass, 
  that would have been the factor.\label{fig:MCorr}}

\figcaption[bulge_ad.ps]{Same as Figure \protect\ref{fig:AdDisk}, 
  but for the bulge shock.
  The adiabatic parameter is given by equation (\protect\ref{eq:adparb}).
  \label{fig:AdBulge}}

\figcaption[evap.ps]{Evaporation time in units of initial relaxation time 
  versus cluster
  concentration. Dots are results of the Fokker-Planck calculations with
  the observational data as input. Solid line - our fit, equation 
  (\protect\ref{eq:fit}).\label{fig:evap}}

\figcaption[trh_c.ps]{Relaxation time at the half-mass radius versus
  cluster concentration for our sample of globular clusters.
  \label{fig:trh}}

\figcaption[ratesr_oc_iso.ps]{Mean probability per Hubble time for cluster 
  destruction due to
  evaporation, and the evaporation plus
  gravitational shocks: disk shock without relaxation, 
  disk shock including relaxation term, disk$+$bulge shocks without relaxation,
  and disk$+$bulge shock including relaxation, as a function of present day
  galactocentric distance. The error bars indicate the interquartile range
  of the distribution. Weinberg adiabatic corrections are used.
  Galactic model is OC, and kinematic model is isotropic.
  \label{fig:ratesr_oc_iso}}

\figcaption[ratesr_oc_ani.ps]{Same as Figure \protect\ref{fig:ratesr_oc_iso}, 
  but for the anisotropic kinematic model.\label{fig:ratesr_oc_ani}}

\figcaption[ratesr_bss_iso.ps]{Same as Figure \protect\ref{fig:ratesr_oc_iso}, 
  but for the BSS galactic model.\label{fig:ratesr_bss_iso}}

\figcaption[ratesr_bss_ani.ps]{Same as Figure \protect\ref{fig:ratesr_oc_ani}, 
  but for the BSS galactic model.\label{fig:ratesr_bss_ani}}

\figcaption[ratesd_oc_iso.ps]{Differential destruction rate for different
  processes: evaporation, disk shock with and without induced relaxation
  (``disk1'' and ``disk'', respectively), and bulge shock with and without
  relaxation term (``bulge1'' and ``bulge''). See text for the definition
  of the differential rate.
  The error bars indicate the interquartile range
  of the distribution. Weinberg adiabatic corrections are used.
  Galactic model is OC, and kinematic model is isotropic.
  \label{fig:ratesd_oc_iso}}

\figcaption[ratesd_oc_ani.ps]{Same as Figure \protect\ref{fig:ratesd_oc_iso}, 
  but for the anisotropic kinematic model.\label{fig:ratesd_oc_ani}}

\figcaption[ratesd_bss_iso.ps]{Same as Figure \protect\ref{fig:ratesd_oc_iso}, 
  but for the BSS galactic model.\label{fig:ratesd_bss_iso}}

\figcaption[ratesd_bss_ani.ps]{Same as Figure \protect\ref{fig:ratesd_oc_ani}, 
  but for the BSS galactic model.\label{fig:ratesd_bss_ani}}

\figcaption[ratesh_oc_iso.ps]{Histogram distribution of the destruction 
  rates from Figure \protect\ref{fig:ratesr_oc_iso}. Weinberg adiabatic 
  correction is used. Galactic model is OC, and kinematic model is isotropic.
  \label{fig:ratesh_oc_iso}}

\figcaption[ratesh_oc_ani.ps]{Histogram distribution of the destruction 
  rates from Figure \protect\ref{fig:ratesr_oc_ani}. 
  Weinberg adiabatic correction is used.
  Galactic model is OC, and kinematic model is anisotropic.
  \label{fig:ratesh_oc_ani}}

\figcaption[ratesh_bss_iso.ps]{Histogram distribution of the destruction 
  rates from Figure \protect\ref{fig:ratesr_bss_iso}. 
  Weinberg adiabatic correction is used.
  Galactic model is BSS, and kinematic model is isotropic.
  \label{fig:ratesh_bss_iso}}

\figcaption[ratesh_bss_ani.ps]{Histogram distribution of the destruction 
  rates from Figure \protect\ref{fig:ratesr_bss_ani}. 
  Weinberg adiabatic correction is used.
  Galactic model is BSS, and kinematic model is anisotropic.
  \label{fig:ratesh_bss_ani}}

\figcaption[diag_oc_iso.ps]{Vital diagram for the Galactic globular clusters.
  Mass-radius plane is restricted by three destructing processes:
  relaxation, tidal shocks, and dynamical friction. Galactic model is OC,
  and the kinematic model is isotropic.\label{fig:diag_oc_iso}}

\figcaption[diag_oc_ani.ps]{Same as Figure \protect\ref{fig:diag_oc_iso},
  but for the anisotropic kinematic model.\label{fig:diag_oc_ani}}

\figcaption[diag_bss_iso.ps]{Same as Figure \protect\ref{fig:diag_oc_iso},
  but for the BSS galactic model.\label{fig:diag_bss_iso}}

\figcaption[diag_bss_ani.ps]{Same as Figure \protect\ref{fig:diag_oc_ani},
  but for the BSS galactic model.\label{fig:diag_bss_ani}}

\figcaption[vital_oc.ps]{Distribution of the destruction times for our sample
  for the OC galactic model. All destruction processes are included.
  Solid line - power-law distribution, normalized to the present time,
  dashes - normalized exponential model.\label{fig:vital_oc}}

\figcaption[vital_bss.ps]{Same as Figure \protect\ref{fig:vital_oc}, but
  for the BSS galactic model.\label{fig:vital_bss}}

\end{document}


\begin{deluxetable}{clrrrrrrrrrr}
\tablecaption{Globular Cluster Parameters}
\tablecolumns{12}
\tablehead{\colhead{ID}  & \colhead{Name} & \colhead{$l$} & \colhead{$b$}   &
           \colhead{$D$} & \colhead{$R$}  & \colhead{$c$} & \colhead{$R_c$} &
           \colhead{$R_t$} & \colhead{$M$} & \colhead{$V_r$} & \colhead{Ref} \\
           & & \colhead{(deg)} & \colhead{(deg)} & 
           \colhead{(kpc)} & \colhead{(kpc)} & & \colhead{(pc)} & 
           \colhead{(pc)} & \colhead{($M_{\sun}$)} & \colhead{(km s$^{-1}$)} &
}
\startdata
\phn\phn 1 &   NGC0104 & 305.90 & -44.89 &   4.6 &   7.8 & 2.04 &  0.50 &  54.8 & 1.45e+06 &  -19.0 &  1\nl
\phn\phn 2 &   NGC0288 & 152.28 & -89.38 &   8.3 &  11.9 & 0.96 &  3.47 &  31.6 & 1.11e+05 &  -46.0 &  1\nl
\phn\phn 3 &   NGC0362 & 301.53 & -46.25 &   8.5 &   9.6 & 1.94 &  0.43 &  37.5 & 3.78e+05 &  223.2 &  1\nl
\phn\phn 4 &   NGC1261 & 270.54 & -52.13 &  15.9 &  18.0 & 1.27 &  1.82 &  33.9 & 3.26e+05 &   51.0 &  9\nl
\phn\phn 5 &      Pal1 & 130.07 &  19.03 &  13.7 &  20.1 & 1.50 &  0.60 &  19.0 & 2.54e+03 &    3.0 &  9\nl
\phn\phn 6 &       AM1 & 258.36 & -48.47 & 124.7 & 126.1 & 1.23 &  4.57 &  77.6 & 1.81e+04 &  116.0 & 11\nl
\phn\phn 7 &  Eridanus & 218.11 & -41.33 &  79.5 &  84.8 & 1.10 &  5.89 &  74.2 & 2.30e+04 &  -21.0 & 10\nl
\phn\phn 8 &      Pal2 & 170.53 &  -9.07 &  13.6 &  21.9 & 1.40 &  0.60 &  15.1 & 1.74e+05 & -133.0 &  9\nl
\phn\phn 9 &   NGC1851 & 244.51 & -35.04 &  12.3 &  17.3 & 2.24 &  0.28 &  48.7 & 5.61e+05 &  320.5 &  1\nl
\phn 10 &   NGC1904 & 227.23 & -29.35 &  12.9 &  19.2 & 1.72 &  0.60 &  31.5 & 3.57e+05 &  203.4 &  1\nl
\phn 11 &   NGC2298 & 245.63 & -16.01 &  10.0 &  15.5 & 1.28 &  1.00 &  19.1 & 7.20e+04 &  150.4 &  3\nl
\phn 12 &   NGC2419 & 180.37 &  25.24 &  83.2 &  91.0 & 1.40 &  8.51 & 213.8 & 1.60e+06 &  -20.0 &  1\nl
\phn 13 &   NGC2808 & 282.19 & -11.25 &   9.1 &  11.1 & 1.77 &  0.71 &  41.8 & 1.32e+06 &   98.0 &  9\nl
\phn 14 &      Pal3 & 240.14 &  41.86 &  90.3 &  93.8 & 1.00 & 12.59 & 125.9 & 6.38e+04 &   84.0 &  9\nl
\phn 15 &   NGC3201 & 277.23 &   8.64 &   5.1 &   9.3 & 1.31 &  2.09 &  42.7 & 1.95e+05 &  494.6 &  1\nl
\phn 16 &      Pal4 & 202.31 &  71.80 &  98.2 & 101.0 & 0.78 & 15.85 &  95.5 & 5.41e+04 &   75.0 &  9\nl
\phn 17 &   NGC4147 & 252.85 &  77.19 &  18.7 &  21.0 & 1.80 &  0.55 &  34.7 & 6.69e+04 &  183.0 &  1\nl
\phn 18 &   NGC4372 & 300.99 &  -9.88 &   5.2 &   7.4 & 1.30 &  2.63 &  52.5 & 3.20e+05 &   49.0 &  9\nl
\phn 19 &    Rup106 & 300.89 &  11.67 &  19.7 &  17.1 & 0.70 &  5.75 &  28.8 & 8.42e+04 &  -44.0 &  7\nl
\phn 20 &   NGC4590 & 299.63 &  36.05 &   9.3 &   9.8 & 1.64 &  1.86 &  81.2 & 3.06e+05 &  -95.1 &  1\nl
\phn 21 &   NGC4833 & 303.61 &  -8.01 &   5.8 &   7.2 & 1.25 &  1.70 &  30.2 & 9.49e+04 &  194.0 &  9\nl
\phn 22 &   NGC5024 & 332.96 &  79.77 &  18.5 &  19.1 & 1.78 &  2.00 & 120.5 & 7.33e+05 &  -80.0 &  9\nl
\phn 23 &   NGC5053 & 335.69 &  78.94 &  16.1 &  16.9 & 0.82 & 10.47 &  69.2 & 1.66e+05 &   42.8 &  1\nl
\phn 24 &   NGC5139 & 309.10 &  14.97 &   4.9 &   6.8 & 1.24 &  3.72 &  64.6 & 2.64e+06 &  232.2 &  1\nl
\phn 25 &   NGC5272 &  42.21 &  78.71 &  10.2 &  12.3 & 1.85 &  1.48 & 104.8 & 7.82e+05 & -146.7 &  1\nl
\phn 26 &   NGC5286 & 311.61 &  10.57 &   9.3 &   7.4 & 1.46 &  0.78 &  22.5 & 4.80e+05 &   57.1 &  1\nl
\phn 27 &   NGC5466 &  42.15 &  73.59 &  15.8 &  16.3 & 1.43 &  8.91 & 239.8 & 1.33e+05 &  107.2 &  1\nl
\phn 28 &   NGC5634 & 342.21 &  49.26 &  25.1 &  20.9 & 1.60 &  1.51 &  60.1 & 2.86e+05 &  -41.9 &  9\nl
\phn 29 &   NGC5694 & 331.06 &  30.36 &  32.8 &  27.0 & 1.84 &  0.59 &  40.8 & 2.92e+05 & -142.8 &  1\nl
\phn 30 &    IC4499 & 307.35 & -20.47 &  18.9 &  15.7 & 1.11 &  5.37 &  69.2 & 1.66e+05 &   41.0 &  6\nl
\phn 31 &   NGC5824 & 332.56 &  22.07 &  32.7 &  26.2 & 2.45 &  0.52 & 146.6 & 8.98e+05 &  -26.1 &  1\nl
\phn 32 &      Pal5 &   0.85 &  45.86 &  23.1 &  18.3 & 0.74 & 19.50 & 107.2 & 2.84e+04 &  -56.0 & 16\nl
\phn 33 &   NGC5897 & 342.95 &  30.29 &  12.5 &   7.3 & 0.79 &  7.24 &  44.6 & 2.00e+05 &  102.9 &  3\nl
\phn 34 &   NGC5904 &   3.86 &  46.80 &   7.5 &   6.5 & 1.87 &  0.89 &  66.0 & 8.34e+05 &   53.9 &  1\nl
\phn 35 &   NGC5927 & 326.60 &   4.86 &   8.0 &   4.8 & 1.60 &  0.98 &  39.0 & 3.71e+05 & -106.0 &  9\nl
\phn 36 &   NGC5946 & 327.58 &   4.19 &  10.4 &   5.6 & 2.50 &  0.23 &  72.7 & 5.66e+05 &  129.0 &  1\nl
\phn 37 &   NGC5986 & 337.02 &  13.27 &  10.2 &   4.6 & 1.22 &  1.91 &  31.7 & 4.98e+05 &   94.0 &  9\nl
\phn 38 &     Pal14 &  28.75 &  42.18 &  73.0 &  67.7 & 0.72 & 23.44 & 123.0 & 2.00e+04 &   72.0 & 10\nl
\phn 39 &   NGC6093 & 352.67 &  19.46 &   8.5 &   3.1 & 1.95 &  0.36 &  32.1 & 3.67e+05 &    7.7 &  1\nl
\phn 40 &   NGC6101 & 317.75 & -15.82 &  14.9 &  10.7 & 0.80 &  5.01 &  31.6 & 1.32e+05 &  364.3 &  3\nl
\phn 41 &   NGC6121 & 350.97 &  15.97 &   2.0 &   6.6 & 1.59 &  0.49 &  19.1 & 2.25e+05 &   70.9 & 14\nl
\phn 42 &   NGC6144 & 351.93 &  15.70 &   9.9 &   3.2 & 1.30 &  2.69 &  53.7 & 1.52e+05 &  162.0 &  9\nl
\phn 43 &   NGC6139 & 342.37 &   6.94 &   9.1 &   3.0 & 1.80 &  0.35 &  22.1 & 4.18e+05 &    4.0 &  9\nl
\phn 44 &   NGC6171 &   3.37 &  23.01 &   6.4 &   3.6 & 1.51 &  1.00 &  32.4 & 2.04e+05 &  -34.0 &  1\nl
\phn 45 &   NGC6205 &  59.01 &  40.91 &   7.1 &   8.7 & 1.49 &  1.82 &  56.2 & 6.27e+05 & -246.4 &  1\nl
\phn 46 &   NGC6218 &  15.72 &  26.31 &   5.5 &   4.7 & 1.38 &  1.07 &  25.7 & 4.93e+05 &  -42.6 &  1\nl
\phn 47 &   NGC6229 &  73.64 &  40.31 &  31.6 &  30.9 & 1.61 &  1.23 &  50.1 & 4.46e+05 & -154.2 &  8\nl
\phn 48 &   NGC6235 & 358.92 &  13.52 &   9.6 &   2.4 & 1.33 &  1.00 &  21.4 & 1.71e+05 &   86.0 &  9\nl
\phn 49 &   NGC6254 &  15.14 &  23.08 &   4.3 &   5.1 & 1.40 &  1.07 &  26.9 & 2.25e+05 &   75.5 &  1\nl
\phn 50 &   NGC6256 & 347.79 &   3.31 &  10.2 &   2.7 & 2.50 &  0.07 &  22.1 & 7.68e+04 & -104.7 &  1\nl
\phn 51 &     Pal15 &  18.87 &  24.30 &  41.3 &  34.2 & 0.60 & 14.45 &  57.5 & 2.74e+04 &   68.9 &  1\nl
\phn 52 &   NGC6266 & 353.57 &   7.32 &   5.6 &   3.1 & 1.70 &  0.29 &  14.5 & 8.98e+05 &  -71.9 &  1\nl
\phn 53 &   NGC6273 & 356.87 &   9.38 &  10.9 &   2.9 & 1.53 &  1.35 &  45.7 & 1.56e+06 &  131.0 &  9\nl
\phn 54 &   NGC6284 & 358.35 &   9.94 &  11.8 &   3.8 & 2.50 &  0.25 &  79.1 & 2.17e+05 &   27.4 &  1\nl
\phn 55 &   NGC6287 &   0.13 &  11.02 &   6.7 &   2.3 & 1.60 &  0.52 &  20.7 & 1.03e+05 & -208.0 &  9\nl
\phn 56 &   NGC6293 & 357.62 &   7.83 &   8.4 &   1.2 & 2.50 &  0.11 &  34.8 & 2.34e+05 & -148.0 &  1\nl
\phn 57 &   NGC6304 & 355.83 &   5.38 &   6.0 &   2.7 & 1.80 &  0.36 &  22.7 & 1.93e+05 &  -98.0 &  8\nl
\phn 58 &   NGC6316 & 357.18 &   5.76 &  12.8 &   4.5 & 1.55 &  0.62 &  22.0 & 1.02e+06 &   76.0 &  9\nl
\phn 59 &   NGC6325 &   0.97 &   8.00 &   6.0 &   2.7 & 2.50 &  0.05 &  15.8 & 9.57e+04 &   30.9 &  1\nl
\phn 60 &   NGC6341 &  68.34 &  34.86 &   7.5 &   9.5 & 1.81 &  0.51 &  32.9 & 3.64e+05 & -120.5 &  8\nl
\phn 61 &   NGC6333 &   5.54 &  10.71 &   7.4 &   2.0 & 1.15 &  1.26 &  17.8 & 2.89e+05 &  224.7 &  8\nl
\phn 62 &   NGC6342 &   4.90 &   9.72 &  12.1 &   4.1 & 2.50 &  0.16 &  50.6 & 1.95e+05 &  117.9 &  1\nl
\phn 63 &   NGC6356 &   6.72 &  10.22 &  16.6 &   8.5 & 1.54 &  1.12 &  38.8 & 8.19e+05 &   31.6 &  8\nl
\phn 64 &   NGC6355 & 359.58 &   5.43 &   7.0 &   1.7 & 2.50 &  0.10 &  31.6 & 3.67e+05 & -184.0 &  9\nl
\phn 65 &   NGC6352 & 341.42 &  -7.17 &   6.1 &   3.5 & 1.10 &  1.48 &  18.6 & 1.22e+05 & -118.0 &  9\nl
\phn 66 &   NGC6366 &  18.41 &  16.04 &   4.0 &   5.1 & 0.92 &  2.14 &  17.8 & 3.92e+04 & -122.6 &  1\nl
\phn 67 &   NGC6362 & 325.56 & -17.57 &   7.8 &   5.4 & 1.10 &  2.95 &  37.1 & 1.17e+05 &  -13.3 &  1\nl
\phn 68 &   NGC6388 & 345.56 &  -6.74 &  11.0 &   3.7 & 1.70 &  0.40 &  20.0 & 1.50e+06 &   77.0 &  9\nl
\phn 69 &   NGC6402 &  21.32 &  14.80 &  10.1 &   4.4 & 1.60 &  2.45 &  97.5 & 1.23e+06 & -123.4 &  8\nl
\phn 70 &   NGC6401 &   3.45 &   3.98 &   6.3 &   2.3 & 1.69 &  0.45 &  22.0 & 1.23e+06 &  -62.0 &  9\nl
\phn 71 &   NGC6397 & 338.17 & -11.96 &   2.2 &   6.5 & 2.50 &  0.03 &   9.5 & 1.59e+05 &   17.7 &  1\nl
\phn 72 &   NGC6426 &  28.09 &  16.23 &  17.5 &  11.3 & 1.70 &  1.35 &  67.7 & 9.49e+04 & -162.0 &  9\nl
\phn 73 &   NGC6440 &   7.73 &   3.80 &   7.1 &   1.8 & 1.70 &  0.26 &  13.0 & 5.72e+05 &  -83.0 &  9\nl
\phn 74 &   NGC6441 & 353.53 &  -5.00 &  10.7 &   2.5 & 1.85 &  0.35 &  24.8 & 1.30e+06 &   16.0 &  1\nl
\phn 75 &   NGC6453 & 355.72 &  -3.87 &  10.1 &   1.8 & 2.50 &  0.20 &  63.2 & 1.32e+05 &  -84.0 &  9\nl
\phn 76 &   NGC6496 & 348.02 & -10.01 &   5.8 &   3.3 & 0.70 &  1.74 &   8.7 & 4.63e+04 &  -95.0 &  9\nl
\phn 77 &   NGC6517 &  19.23 &   6.76 &   6.1 &   3.5 & 1.82 &  0.11 &   7.3 & 1.82e+05 &  -37.0 &  9\nl
\phn 78 &   NGC6522 &   1.02 &  -3.93 &   7.2 &   1.4 & 2.50 &  0.11 &  34.8 & 5.93e+04 &  -10.4 &  1\nl
\phn 79 &   NGC6535 &  27.18 &  10.44 &   7.0 &   4.1 & 1.30 &  0.83 &  16.6 & 5.93e+04 & -215.3 &  1\nl
\phn 80 &   NGC6528 &   1.14 &  -4.17 &   6.6 &   2.0 & 2.29 &  0.17 &  33.1 & 9.31e+04 &  160.0 &  9\nl
\phn 81 &   NGC6539 &  20.80 &   6.78 &   4.0 &   5.0 & 1.60 &  0.63 &  25.1 & 1.60e+05 &  -52.0 &  9\nl
\phn 82 &   NGC6544 &   5.84 &  -2.20 &   2.6 &   5.9 & 2.50 &  0.03 &   9.5 & 1.30e+05 &  -12.0 &  8\nl
\phn 83 &   NGC6541 & 349.29 & -11.18 &   6.7 &   2.7 & 2.00 &  0.58 &  58.0 & 4.67e+05 & -152.8 &  8\nl
\phn 84 &   NGC6553 &   5.25 &  -3.02 &   3.6 &   4.9 & 1.17 &  0.56 &   8.3 & 2.61e+05 &  -24.0 &  9\nl
\phn 85 &   NGC6558 &   0.20 &  -6.03 &   8.9 &   1.0 & 2.50 &  0.09 &  28.5 & 2.41e+05 & -198.9 &  1\nl
\phn 86 &   NGC6569 &   0.48 &  -6.68 &   8.8 &   1.0 & 1.27 &  0.95 &  17.7 & 3.85e+05 &  -26.0 &  9\nl
\phn 87 &   NGC6584 & 342.14 & -16.41 &   9.5 &   3.9 & 1.20 &  1.66 &  26.3 & 2.19e+05 &  180.0 &  8\nl
\phn 88 &   NGC6624 &   2.79 &  -7.91 &   8.1 &   1.3 & 2.50 &  0.15 &  47.4 & 3.32e+05 &   54.2 &  1\nl
\phn 89 &   NGC6626 &   7.80 &  -5.58 &   6.0 &   2.8 & 1.67 &  0.42 &  19.6 & 4.42e+05 &   15.5 &  1\nl
\phn 90 &   NGC6638 &   7.90 &  -7.15 &   8.4 &   1.6 & 1.40 &  0.65 &  16.3 & 1.12e+05 &  -14.0 &  8\nl
\phn 91 &   NGC6637 &   1.72 & -10.27 &  10.2 &   2.4 & 1.39 &  1.00 &  24.5 & 3.44e+05 &   50.1 &  8\nl
\phn 92 &   NGC6642 &   9.81 &  -6.44 &   7.9 &   1.8 & 1.99 &  0.24 &  23.5 & 1.26e+05 &  -47.0 &  9\nl
\phn 93 &   NGC6652 &   1.53 & -11.38 &  14.3 &   6.2 & 1.80 &  0.30 &  18.9 & 2.59e+05 & -124.2 &  8\nl
\phn 94 &   NGC6656 &   9.89 &  -7.55 &   3.0 &   5.6 & 1.31 &  1.23 &  25.1 & 5.36e+05 & -148.8 & 13\nl
\phn 95 &      Pal8 &  14.11 &  -6.80 &  28.6 &  20.6 & 0.75 &  3.31 &  18.6 & 2.21e+05 &  -38.0 & 15\nl
\phn 96 &   NGC6681 &   2.85 & -12.51 &   9.2 &   2.1 & 2.50 &  0.07 &  22.1 & 1.89e+05 &  221.1 &  1\nl
\phn 97 &   NGC6712 &  25.35 &  -4.32 &   6.8 &   3.8 & 0.90 &  1.86 &  14.8 & 2.45e+05 & -107.5 &  1\nl
\phn 98 &   NGC6715 &   5.61 & -14.09 &  21.1 &  13.1 & 1.84 &  0.66 &  45.7 & 1.45e+06 &  142.0 &  3\nl
\phn 99 &   NGC6717 &  12.88 & -10.90 &   7.3 &   2.6 & 2.07 &  0.18 &  21.1 & 1.23e+05 &   -6.0 &  9\nl
100 &   NGC6723 &   0.07 & -17.30 &   8.6 &   2.6 & 1.05 &  2.40 &  26.9 & 3.74e+05 &  -79.0 &  9\nl
101 &   NGC6752 & 336.50 & -25.63 &   4.2 &   5.5 & 2.50 &  0.21 &  66.4 & 3.64e+05 &  -32.1 &  1\nl
102 &   NGC6760 &  36.11 &  -3.92 &   6.1 &   5.1 & 1.59 &  0.59 &  23.0 & 1.38e+05 &  -28.0 &  9\nl
103 &   Terzan7 &   3.39 & -20.07 &  36.7 &  28.9 & 1.08 &  6.46 &  77.7 & 6.15e+04 &  162.0 &  2\nl
104 &   NGC6779 &  62.66 &   8.34 &   9.7 &   9.6 & 1.37 &  1.05 &  24.6 & 1.89e+05 & -135.9 &  1\nl
105 &      Arp2 &   8.55 & -20.79 &  28.2 &  20.6 & 0.90 & 13.18 & 104.7 & 3.26e+04 &  119.0 &  2\nl
106 &   NGC6809 &   8.79 & -23.27 &   4.8 &   4.6 & 0.76 &  3.98 &  22.9 & 2.41e+05 &  175.5 &  1\nl
107 &   Terzan8 &   5.76 & -24.56 &  23.0 &  15.7 & 0.60 &  6.61 &  26.3 & 2.50e+04 &  130.0 &  2\nl
108 &     Pal11 &  31.80 & -15.58 &  13.2 &   7.9 & 0.69 &  7.76 &  38.0 & 1.38e+05 &  -68.0 &  8\nl
109 &   NGC6838 &  56.74 &  -4.56 &   3.9 &   7.2 & 1.15 &  0.72 &  10.2 & 3.67e+04 &  -23.0 &  1\nl
110 &   NGC6864 &  20.30 & -25.75 &  18.7 &  12.4 & 1.88 &  0.51 &  38.7 & 4.89e+05 & -195.2 &  8\nl
111 &   NGC6934 &  52.10 & -18.89 &  14.8 &  12.0 & 1.53 &  1.07 &  36.3 & 2.15e+05 & -412.2 &  1\nl
112 &   NGC6981 &  35.16 & -32.68 &  18.1 &  13.7 & 1.23 &  2.82 &  47.9 & 1.81e+05 & -309.0 &  9\nl
113 &   NGC7006 &  63.77 & -19.41 &  38.8 &  36.1 & 1.42 &  2.75 &  72.3 & 2.52e+05 & -364.0 &  9\nl
114 &   NGC7078 &  65.01 & -27.31 &  10.4 &  10.7 & 2.50 &  0.21 &  66.4 & 9.84e+05 & -107.1 & 12\nl
115 &   NGC7089 &  53.38 & -35.78 &  11.8 &  10.7 & 1.80 &  1.17 &  73.8 & 8.81e+05 &   -3.1 &  1\nl
116 &   NGC7099 &  27.18 & -46.84 &   7.4 &   7.1 & 2.50 &  0.12 &  37.9 & 2.74e+05 & -185.7 &  1\nl
117 &     Pal12 &  30.51 & -47.68 &  19.1 &  15.8 & 0.90 &  6.17 &  49.0 & 2.02e+04 &   28.5 &  5\nl
118 &     Pal13 &  87.10 & -42.70 &  24.5 &  25.6 & 1.00 &  2.82 &  28.2 & 5.12e+03 &   13.0 &  4\nl
119 &   NGC7492 &  53.39 & -63.48 &  25.0 &  24.2 & 1.00 &  6.17 &  61.7 & 5.56e+04 & -188.5 &  8\nl
\enddata
\tablerefs{
(1) Pryor \& Meylan 1993 (in case of multiple choice, mean values are assumed);
(2) Da Costa \& Armandroff 1995;
(3) Geisler et al. 1995;
(4) Kulessa \& McDowell 1985;
(5) Armandroff \& Da Costa 1991;
(6) Cannon 1993;
(7) Da Costa, Armandroff, \& Norris 1992;
(8) Webbink 1981;
(9) Hesser, Shawl, \& Meyer 1986;
(10) Zaritsky et al. 1989;
(11) Suntzeff, Olszewski, \& Stetson 1985;
(12) Peterson, Seitzer, \& Cudworth 1989;
(13) Peterson, \& Cudworth 1994;
(14) Peterson, Rees, \& Cudworth 1995;
(15) Zinn \& West 1984;
(16) Schweitzer, Cudworth, \& Majewski 1993.}
\end{deluxetable}


\begin{deluxetable}{clrrrrrrrr}
\tablecaption{Destruction Rates for Clusters in Sample\tablenotemark{a}}
\tablecolumns{10}
\setcounter{table}{2}
\tablehead{& & & & \multicolumn{3}{c}{OC isotropic} & 
           \multicolumn{3}{c}{BSS isotropic} \\ \cline{5-7} \cline{8-10}
           \colhead{ID} & \colhead{Name} & \colhead{$t_{rh}$ (yr)} & 
           \colhead{$\nu_{evap}$} & \colhead{$\nu_{d1+d2}$} & 
           \colhead{$\nu_{d1+b1}$} & \colhead{$\nu_{tot}$} &
           \colhead{$\nu_{d1+d2}$} & \colhead{$\nu_{d1+b1}$} &
           \colhead{$\nu_{tot}$}}
\startdata
\phn\phn 1 &   NGC0104 & 3.94e+09 & 7.10e-02 & 7.84e-02 & 7.36e-02 & 7.91e-02 & 1.37e-01 & 1.11e-01 & 1.49e-01\nl
\phn\phn 2 &   NGC0288 & 1.42e+09 & 3.71e-01 & 4.03e-01 & 4.47e+00 & 1.93e+01 & 8.80e-01 & 6.37e-01 & 1.10e+00\nl
\phn\phn 3 &   NGC0362 & 1.18e+09 & 2.18e-01 & 2.28e-01 & 2.24e-01 & 2.33e-01 & 3.34e-01 & 2.85e-01 & 3.54e-01\nl
\phn\phn 4 &   NGC1261 & 1.51e+09 & 2.33e-01 & 2.39e-01 & 2.39e-01 & 2.48e-01 & 2.74e-01 & 2.50e-01 & 2.79e-01\nl
\phn\phn 5 &      Pal1 & 6.87e+07 & 4.37e+00 & 5.34e+00 & 1.39e+01 & 2.26e+01 & 4.47e+00 & 4.47e+00 & 4.49e+00\nl
\phn\phn 6 &       AM1 & 1.72e+09 & 2.20e-01 & 4.03e-01 & 3.80e-01 & 6.02e-01 & 2.20e-01 & 2.20e-01 & 2.20e-01\nl
\phn\phn 7 &  Eridanus & 2.16e+09 & 2.03e-01 & 2.07e-01 & 2.04e-01 & 2.06e-01 & 2.03e-01 & 2.03e-01 & 2.03e-01\nl
\phn\phn 8 &      Pal2 & 2.87e+08 & 1.06e+00 & 1.06e+00 & 1.09e+00 & 1.08e+00 & 1.06e+00 & 1.06e+00 & 1.06e+00\nl
\phn\phn 9 &   NGC1851 & 2.72e+09 & 1.24e-01 & 1.27e-01 & 1.25e-01 & 1.27e-01 & 1.28e-01 & 1.26e-01 & 1.24e-01\nl
\phn 10 &   NGC1904 & 8.77e+08 & 2.83e-01 & 2.85e-01 & 2.84e-01 & 2.84e-01 & 3.40e-01 & 3.10e-01 & 3.54e-01\nl
\phn 11 &   NGC2298 & 3.36e+08 & 1.02e+00 & 1.02e+00 & 1.02e+00 & 1.05e+00 & 1.03e+00 & 1.02e+00 & 1.03e+00\nl
\phn 12 &   NGC2419 & 3.90e+10 & 7.97e-03 & 8.00e-03 & 7.98e-03 & 7.99e-03 & 7.97e-03 & 8.03e-03 & 8.01e-03\nl
\phn 13 &   NGC2808 & 2.30e+09 & 1.16e-01 & 1.10e-01 & 1.10e-01 & 1.11e-01 & 1.50e-01 & 1.32e-01 & 1.61e-01\nl
\phn 14 &      Pal3 & 8.46e+09 & 5.88e-02 & 5.88e-02 & 5.88e-02 & 5.88e-02 & 5.88e-02 & 5.88e-02 & 5.88e-02\nl
\phn 15 &   NGC3201 & 1.62e+09 & 2.07e-01 & 2.63e-01 & 2.28e-01 & 2.69e-01 & 3.30e-01 & 2.65e-01 & 3.45e-01\nl
\phn 16 &      Pal4 & 7.38e+09 & 1.16e-01 & 1.16e-01 & 1.16e-01 & 1.16e-01 & 1.16e-01 & 1.16e-01 & 1.16e-01\nl
\phn 17 &   NGC4147 & 5.00e+08 & 4.83e-01 & 4.84e-01 & 4.84e-01 & 4.84e-01 & 4.83e-01 & 4.83e-01 & 4.83e-01\nl
\phn 18 &   NGC4372 & 2.76e+09 & 1.24e-01 & 1.48e-01 & 1.36e-01 & 1.58e-01 & 4.59e-01 & 3.04e-01 & 4.74e-01\nl
\phn 19 &    Rup106 & 1.64e+09 & 5.03e+00 & 7.88e+00 & 6.00e+00 & 7.97e+00 & 5.11e+00 & 5.05e+00 & 5.13e+00\nl
\phn 20 &   NGC4590 & 3.56e+09 & 7.17e-02 & 1.08e-01 & 8.16e-02 & 9.82e-02 & 8.12e-02 & 7.55e-02 & 8.22e-02\nl
\phn 21 &   NGC4833 & 7.86e+08 & 4.52e-01 & 4.83e-01 & 5.91e-01 & 7.65e-01 & 1.60e+00 & 2.40e+00 & 4.36e+00\nl
\phn 22 &   NGC5024 & 8.74e+09 & 2.88e-02 & 2.84e-02 & 2.88e-02 & 2.88e-02 & 3.03e-02 & 2.96e-02 & 3.07e-02\nl
\phn 23 &   NGC5053 & 6.76e+09 & 1.07e-01 & 2.46e+00 & 4.81e+00 & 1.39e+01 & 1.36e+00 & 2.40e-01 & 1.62e+00\nl
\phn 24 &   NGC5139 & 1.01e+10 & 3.65e-02 & 6.58e-02 & 1.01e-01 & 1.60e-01 & 1.49e-01 & 1.92e-01 & 3.43e-01\nl
\phn 25 &   NGC5272 & 7.28e+09 & 3.41e-02 & 4.01e-02 & 3.66e-02 & 3.98e-02 & 5.19e-02 & 4.29e-02 & 5.42e-02\nl
\phn 26 &   NGC5286 & 7.42e+08 & 3.97e-01 & 4.00e-01 & 3.96e-01 & 4.01e-01 & 4.57e-01 & 1.07e+00 & 1.57e+00\nl
\phn 27 &   NGC5466 & 1.57e+10 & 1.88e-02 & 1.02e+00 & 4.69e-01 & 3.01e+00 & 1.66e+00 & 4.30e-01 & 2.73e+00\nl
\phn 28 &   NGC5634 & 2.27e+09 & 1.18e-01 & 1.18e-01 & 1.17e-01 & 1.19e-01 & 1.15e-01 & 1.18e-01 & 1.15e-01\nl
\phn 29 &   NGC5694 & 1.17e+09 & 2.13e-01 & 2.11e-01 & 2.17e-01 & 2.13e-01 & 2.10e-01 & 2.10e-01 & 2.10e-01\nl
\phn 30 &    IC4499 & 4.26e+09 & 9.93e-02 & 1.07e-01 & 1.02e-01 & 1.11e-01 & 4.57e-01 & 2.34e-01 & 4.76e-01\nl
\phn 31 &   NGC5824 & 2.13e+10 & 1.96e-02 & 2.08e-02 & 2.02e-02 & 2.10e-02 & 2.71e-02 & 2.41e-02 & 3.06e-02\nl
\phn 32 &      Pal5 & 7.17e+09 & 1.64e-01 & 2.37e+00 & 3.79e+00 & 9.62e+00 & 1.74e-01 & 1.66e-01 & 2.16e-01\nl
\phn 33 &   NGC5897 & 3.96e+09 & 2.05e-01 & 1.49e+00 & 6.55e+00 & 1.85e+01 & 3.71e+01 & 1.63e+02 & 4.18e+02\nl
\phn 34 &   NGC5904 & 3.75e+09 & 6.76e-02 & 8.61e-02 & 8.57e-02 & 1.08e-01 & 4.06e-01 & 1.11e+00 & 1.76e+00\nl
\phn 35 &   NGC5927 & 1.32e+09 & 1.98e-01 & 2.23e-01 & 2.14e-01 & 2.43e-01 & 3.57e-01 & 2.71e-01 & 3.49e-01\nl
\phn 36 &   NGC5946 & 6.36e+09 & 6.80e-02 & 1.73e-01 & 2.17e-01 & 3.43e-01 & 2.32e-01 & 1.46e-01 & 2.38e-01\nl
\phn 37 &   NGC5986 & 1.76e+09 & 2.13e-01 & 2.20e-01 & 9.37e-01 & 1.74e+00 & 5.42e-01 & 1.81e+00 & 9.70e+00\nl
\phn 38 &     Pal14 & 7.90e+09 & 1.78e-01 & 1.78e-01 & 1.83e-01 & 1.96e-01 & 3.97e+00 & 9.41e+00 & 2.05e+01\nl
\phn 39 &   NGC6093 & 9.26e+08 & 2.77e-01 & 2.84e-01 & 2.88e+00 & 4.20e+00 & 7.68e-01 & 1.35e+00 & 1.86e+00\nl
\phn 40 &   NGC6101 & 1.96e+09 & 3.97e-01 & 4.05e-01 & 4.00e-01 & 4.05e-01 & 9.38e+00 & 1.32e+01 & 3.76e+01\nl
\phn 41 &   NGC6121 & 3.70e+08 & 7.05e-01 & 7.16e-01 & 7.54e-01 & 8.30e-01 & 7.38e-01 & 7.37e-01 & 7.52e-01\nl
\phn 42 &   NGC6144 & 2.10e+09 & 1.65e-01 & 3.26e-01 & 4.78e-01 & 7.21e-01 & 1.71e+00 & 9.30e-01 & 2.00e+00\nl
\phn 43 &   NGC6139 & 5.40e+08 & 4.53e-01 & 4.56e-01 & 4.55e-01 & 4.61e-01 & 7.13e-01 & 2.16e+00 & 3.07e+00\nl
\phn 44 &   NGC6171 & 8.48e+08 & 3.26e-01 & 3.57e-01 & 3.40e-01 & 3.77e-01 & 5.59e-01 & 4.22e-01 & 5.48e-01\nl
\phn 45 &   NGC6205 & 3.17e+09 & 9.13e-02 & 1.23e-01 & 1.15e-01 & 1.47e-01 & 1.01e-01 & 9.55e-02 & 1.02e-01\nl
\phn 46 &   NGC6218 & 1.01e+09 & 3.13e-01 & 3.16e-01 & 3.87e-01 & 4.68e-01 & 3.78e-01 & 3.39e-01 & 3.84e-01\nl
\phn 47 &   NGC6229 & 2.06e+09 & 1.41e-01 & 1.28e-01 & 1.27e-01 & 1.28e-01 & 1.27e-01 & 1.28e-01 & 1.31e-01\nl
\phn 48 &   NGC6235 & 5.30e+08 & 6.18e-01 & 6.20e-01 & 6.78e+00 & 3.26e+01 & 9.06e-01 & 7.99e-01 & 1.02e+00\nl
\phn 49 &   NGC6254 & 7.60e+08 & 4.01e-01 & 4.33e-01 & 4.51e-01 & 5.03e-01 & 4.22e-01 & 4.08e-01 & 4.23e-01\nl
\phn 50 &   NGC6256 & 4.67e+08 & 8.68e-01 & 8.95e-01 & 9.72e-01 & 1.08e+00 & 1.76e+00 & 1.29e+00 & 1.78e+00\nl
\phn 51 &     Pal15 & 3.33e+09 & 3.80e+02 & 3.80e+02 & 3.80e+02 & 3.80e+02 & 3.81e+02 & 3.81e+02 & 3.83e+02\nl
\phn 52 &   NGC6266 & 4.09e+08 & 6.27e-01 & 6.28e-01 & 6.33e-01 & 6.44e-01 & 6.26e-01 & 6.31e-01 & 6.38e-01\nl
\phn 53 &   NGC6273 & 3.29e+09 & 9.68e-02 & 9.24e-02 & 1.02e-01 & 1.20e-01 & 2.89e-01 & 4.72e+00 & 2.46e+01\nl
\phn 54 &   NGC6284 & 4.82e+09 & 8.68e-02 & 4.69e-01 & 3.79e-01 & 5.98e-01 & 2.26e+00 & 3.11e+00 & 5.34e+00\nl
\phn 55 &   NGC6287 & 3.01e+08 & 8.88e-01 & 8.70e-01 & 1.52e+00 & 2.09e+00 & 9.78e-01 & 9.62e-01 & 1.13e+00\nl
\phn 56 &   NGC6293 & 1.45e+09 & 2.88e-01 & 3.02e-01 & 6.88e+00 & 9.66e+00 & 1.47e+00 & 6.21e+00 & 9.05e+00\nl
\phn 57 &   NGC6304 & 4.08e+08 & 5.94e-01 & 6.08e-01 & 2.91e+00 & 4.37e+00 & 6.60e-01 & 6.22e-01 & 6.74e-01\nl
\phn 58 &   NGC6316 & 8.97e+08 & 3.10e-01 & 3.06e-01 & 6.71e-01 & 9.61e-01 & 3.31e-01 & 3.09e-01 & 3.13e-01\nl
\phn 59 &   NGC6325 & 3.08e+08 & 1.33e+00 & 1.35e+00 & 2.52e+00 & 3.54e+00 & 1.66e+00 & 1.45e+00 & 1.67e+00\nl
\phn 60 &   NGC6341 & 9.27e+08 & 2.65e-01 & 2.68e-01 & 2.64e-01 & 2.66e-01 & 3.60e-01 & 3.25e-01 & 4.01e-01\nl
\phn 61 &   NGC6333 & 6.57e+08 & 6.15e-01 & 6.18e-01 & 6.19e-01 & 6.26e-01 & 7.24e-01 & 1.82e+00 & 3.11e+00\nl
\phn 62 &   NGC6342 & 2.36e+09 & 1.80e-01 & 3.23e-01 & 3.09e-01 & 4.53e-01 & 1.66e+00 & 3.55e+00 & 5.27e+00\nl
\phn 63 &   NGC6356 & 1.94e+09 & 1.44e-01 & 1.43e-01 & 1.44e-01 & 1.43e-01 & 1.82e-01 & 1.59e-01 & 1.83e-01\nl
\phn 64 &   NGC6355 & 1.52e+09 & 2.79e-01 & 2.96e-01 & 3.78e-01 & 4.89e-01 & 9.20e-01 & 1.59e+00 & 2.47e+00\nl
\phn 65 &   NGC6352 & 5.33e+08 & 8.02e-01 & 8.14e-01 & 8.06e-01 & 8.14e-01 & 1.04e+00 & 8.93e-01 & 1.05e+00\nl
\phn 66 &   NGC6366 & 4.20e+08 & 1.37e+00 & 1.41e+00 & 1.45e+00 & 1.63e+00 & 3.25e+00 & 3.13e+00 & 8.64e+00\nl
\phn 67 &   NGC6362 & 1.48e+09 & 2.90e-01 & 4.26e-01 & 3.64e-01 & 4.78e-01 & 7.29e-01 & 4.60e-01 & 7.35e-01\nl
\phn 68 &   NGC6388 & 8.25e+08 & 3.08e-01 & 3.07e-01 & 3.08e-01 & 3.16e-01 & 3.22e-01 & 3.18e-01 & 3.53e-01\nl
\phn 69 &   NGC6402 & 8.68e+09 & 3.08e-02 & 7.99e-02 & 4.72e+00 & 2.06e+01 & 2.65e-01 & 2.17e-01 & 3.32e-01\nl
\phn 70 &   NGC6401 & 8.78e+08 & 2.89e-01 & 2.90e-01 & 7.69e-01 & 1.09e+00 & 2.97e-01 & 3.02e-01 & 3.00e-01\nl
\phn 71 &   NGC6397 & 1.76e+08 & 2.35e+00 & 2.36e+00 & 2.36e+00 & 2.42e+00 & 2.43e+00 & 2.37e+00 & 2.48e+00\nl
\phn 72 &   NGC6426 & 1.61e+09 & 1.51e-01 & 2.54e-01 & 2.25e-01 & 3.14e-01 & 1.77e-01 & 1.63e-01 & 1.86e-01\nl
\phn 73 &   NGC6440 & 2.87e+08 & 8.77e-01 & 8.78e-01 & 8.78e-01 & 8.89e-01 & 9.01e-01 & 8.98e-01 & 9.05e-01\nl
\phn 74 &   NGC6441 & 1.04e+09 & 2.64e-01 & 2.42e-01 & 3.39e-01 & 4.28e-01 & 2.74e-01 & 2.57e-01 & 2.86e-01\nl
\phn 75 &   NGC6453 & 2.81e+09 & 1.51e-01 & 8.28e-01 & 8.88e-01 & 1.38e+00 & 3.51e+00 & 2.40e+00 & 3.66e+00\nl
\phn 76 &   NGC6496 & 2.14e+08 & 3.85e+01 & 3.86e+01 & 1.69e+02 & 3.36e+02 & 5.97e+01 & 1.55e+02 & 3.05e+02\nl
\phn 77 &   NGC6517 & 7.21e+07 & 3.37e+00 & 3.37e+00 & 3.50e+00 & 3.46e+00 & 3.46e+00 & 3.37e+00 & 3.40e+00\nl
\phn 78 &   NGC6522 & 8.28e+08 & 4.98e-01 & 8.31e-01 & 3.24e+00 & 5.04e+00 & 5.23e+00 & 6.08e+00 & 9.18e+00\nl
\phn 79 &   NGC6535 & 2.45e+08 & 1.37e+00 & 1.40e+00 & 1.54e+00 & 1.79e+00 & 1.81e+00 & 1.66e+00 & 1.97e+00\nl
\phn 80 &   NGC6528 & 7.65e+08 & 4.57e-01 & 6.90e-01 & 1.99e+00 & 3.00e+00 & 2.65e+00 & 2.58e+00 & 3.89e+00\nl
\phn 81 &   NGC6539 & 4.81e+08 & 5.47e-01 & 5.44e-01 & 5.37e-01 & 5.49e-01 & 5.83e-01 & 5.54e-01 & 5.92e-01\nl
\phn 82 &   NGC6544 & 1.62e+08 & 2.55e+00 & 2.55e+00 & 2.55e+00 & 2.55e+00 & 2.55e+00 & 2.55e+00 & 2.55e+00\nl
\phn 83 &   NGC6541 & 2.57e+09 & 1.05e-01 & 1.80e-01 & 1.74e-01 & 2.42e-01 & 5.31e-01 & 4.50e-01 & 6.42e-01\nl
\phn 84 &   NGC6553 & 1.94e+08 & 2.04e+00 & 2.04e+00 & 2.04e+00 & 2.10e+00 & 2.05e+00 & 2.04e+00 & 2.05e+00\nl
\phn 85 &   NGC6558 & 1.09e+09 & 3.85e-01 & 3.93e-01 & 1.52e+00 & 2.41e+00 & 2.56e+00 & 3.74e+01 & 5.57e+01\nl
\phn 86 &   NGC6569 & 6.09e+08 & 5.82e-01 & 5.80e-01 & 1.35e+00 & 2.01e+00 & 9.62e-01 & 5.96e+00 & 3.05e+01\nl
\phn 87 &   NGC6584 & 9.72e+08 & 4.02e-01 & 4.03e-01 & 4.41e-01 & 5.18e-01 & 6.82e-01 & 6.28e-01 & 8.54e-01\nl
\phn 88 &   NGC6624 & 2.68e+09 & 1.58e-01 & 1.62e-01 & 1.32e+00 & 1.83e+00 & 1.86e+00 & 2.79e+00 & 3.94e+00\nl
\phn 89 &   NGC6626 & 4.84e+08 & 5.22e-01 & 5.28e-01 & 5.29e-01 & 5.46e-01 & 6.96e-01 & 1.09e+00 & 1.49e+00\nl
\phn 90 &   NGC6638 & 2.70e+08 & 1.12e+00 & 1.13e+00 & 8.22e+00 & 1.41e+01 & 1.70e+00 & 3.07e+00 & 4.39e+00\nl
\phn 91 &   NGC6637 & 8.03e+08 & 3.88e-01 & 3.91e-01 & 1.21e+00 & 1.81e+00 & 5.37e-01 & 4.63e-01 & 5.45e-01\nl
\phn 92 &   NGC6642 & 3.81e+08 & 6.69e-01 & 7.03e-01 & 1.38e+00 & 1.90e+00 & 1.88e+00 & 4.03e+00 & 5.56e+00\nl
\phn 93 &   NGC6652 & 3.51e+08 & 7.08e-01 & 6.95e-01 & 7.08e-01 & 7.28e-01 & 6.96e-01 & 6.94e-01 & 6.95e-01\nl
\phn 94 &   NGC6656 & 1.12e+09 & 3.03e-01 & 3.11e-01 & 3.06e-01 & 3.15e-01 & 3.24e-01 & 3.09e-01 & 3.21e-01\nl
\phn 95 &      Pal8 & 1.18e+09 & 8.70e-01 & 8.72e-01 & 8.71e-01 & 8.72e-01 & 8.70e-01 & 8.70e-01 & 8.70e-01\nl
\phn 96 &   NGC6681 & 6.75e+08 & 6.34e-01 & 6.26e-01 & 6.51e-01 & 7.09e-01 & 8.55e-01 & 1.06e+00 & 1.48e+00\nl
\phn 97 &   NGC6712 & 6.94e+08 & 8.54e-01 & 8.57e-01 & 1.03e+00 & 1.35e+00 & 1.16e+00 & 7.46e+00 & 3.59e+01\nl
\phn 98 &   NGC6715 & 2.72e+09 & 9.24e-02 & 9.41e-02 & 9.41e-02 & 9.56e-02 & 1.83e-01 & 2.90e-01 & 4.03e-01\nl
\phn 99 &   NGC6717 & 3.44e+08 & 8.15e-01 & 7.99e-01 & 6.04e+00 & 8.97e+00 & 1.80e+00 & 4.30e+00 & 6.13e+00\nl
100 &   NGC6723 & 1.60e+09 & 2.93e-01 & 2.97e-01 & 2.96e-01 & 3.21e-01 & 5.04e-01 & 4.09e-01 & 5.77e-01\nl
101 &   NGC6752 & 4.61e+09 & 9.19e-02 & 2.40e-01 & 1.60e-01 & 2.44e-01 & 3.08e-01 & 2.04e-01 & 3.23e-01\nl
102 &   NGC6760 & 3.99e+08 & 6.45e-01 & 6.87e-01 & 6.69e-01 & 7.16e-01 & 9.49e-01 & 8.86e-01 & 1.11e+00\nl
103 &   Terzan7 & 3.55e+09 & 1.24e-01 & 1.26e-01 & 1.29e-01 & 1.28e-01 & 1.28e-01 & 1.25e-01 & 1.25e-01\nl
104 &   NGC6779 & 6.46e+08 & 4.86e-01 & 4.90e-01 & 4.88e-01 & 5.01e-01 & 5.21e-01 & 5.12e-01 & 5.24e-01\nl
105 &      Arp2 & 5.75e+09 & 1.04e-01 & 1.21e-01 & 1.22e-01 & 1.61e-01 & 3.59e+01 & 1.94e+01 & 6.02e+01\nl
106 &   NGC6809 & 1.64e+09 & 5.84e-01 & 6.03e-01 & 6.42e-01 & 7.52e-01 & 6.67e-01 & 6.06e-01 & 6.83e-01\nl
107 &   Terzan8 & 9.94e+08 & 1.27e+03 & 1.28e+03 & 1.28e+03 & 1.28e+03 & 1.28e+03 & 1.28e+03 & 1.28e+03\nl
108 &     Pal11 & 3.07e+09 & 7.62e+00 & 1.21e+01 & 9.16e+00 & 1.21e+01 & 1.36e+01 & 9.85e+00 & 1.38e+01\nl
109 &   NGC6838 & 1.22e+08 & 3.40e+00 & 3.33e+00 & 3.41e+00 & 3.34e+00 & 3.34e+00 & 3.35e+00 & 3.34e+00\nl
110 &   NGC6864 & 1.35e+09 & 1.87e-01 & 1.91e-01 & 1.88e-01 & 1.92e-01 & 1.91e-01 & 1.88e-01 & 1.89e-01\nl
111 &   NGC6934 & 1.01e+09 & 2.80e-01 & 2.74e-01 & 2.74e-01 & 2.75e-01 & 2.85e-01 & 2.77e-01 & 2.89e-01\nl
112 &   NGC6981 & 2.11e+09 & 1.73e-01 & 1.79e-01 & 1.74e-01 & 1.76e-01 & 1.83e-01 & 1.74e-01 & 1.76e-01\nl
113 &   NGC7006 & 3.43e+09 & 8.72e-02 & 8.96e-02 & 8.83e-02 & 9.02e-02 & 8.72e-02 & 8.73e-02 & 8.72e-02\nl
114 &   NGC7078 & 7.01e+09 & 6.23e-02 & 1.18e-01 & 1.35e-01 & 2.03e-01 & 2.02e-01 & 1.38e-01 & 2.17e-01\nl
115 &   NGC7089 & 4.52e+09 & 5.51e-02 & 7.59e-02 & 6.74e-02 & 8.13e-02 & 5.85e-02 & 5.66e-02 & 5.76e-02\nl
116 &   NGC7099 & 1.77e+09 & 2.38e-01 & 2.70e-01 & 2.55e-01 & 2.88e-01 & 4.65e-01 & 4.04e-01 & 5.65e-01\nl
117 &     Pal12 & 1.52e+09 & 3.98e-01 & 4.12e-01 & 4.12e-01 & 4.15e-01 & 3.79e+00 & 1.06e+00 & 7.12e+00\nl
118 &     Pal13 & 3.34e+08 & 1.57e+00 & 1.57e+00 & 1.57e+00 & 1.57e+00 & 1.57e+00 & 1.57e+00 & 1.57e+00\nl
119 &   NGC7492 & 2.75e+09 & 1.85e-01 & 2.24e-01 & 2.28e-01 & 3.16e-01 & 4.72e+00 & 3.79e+00 & 1.44e+01\nl
\enddata
\tablenotetext{a}{Destruction rates $\nu$ in units of inverse Hubble time.}
\tablecomments{The rates including the tidal shocks are: $\nu_{d1+d2}$ --
first and second order disk shock, $\nu_{d1+b1}$ -- first order disk and bulge
shocks, $\nu_{tot}$ -- total destruction rate (first and second order 
disk+bulge shocks). The two groups of columns correspond to the results for 
the OC and BSS galactic models. In both cases the isotropic kinematic model 
is assumed, appropriate for the present clusters. The Weinberg's form of the 
adiabatic correction is used (see text for more detail).}
\end{deluxetable}